\documentstyle[11pt,aaspp]{article}
\input{psfig}
\eqsecnum
\def\beq{\begin{equation}}
\def\eeq{\end{equation}}

\begin{document}

\title{Formation of Millisecond Pulsars from Accretion Induced Collapse and
Constraints on Pulsar Gamma Ray Burst Models}
\author{Insu Yi$^1$ and Eric G. Blackman$^2$}
\affil{$^1$Institute for Advanced Study, Olden Lane, Princeton, NJ 08540; 
yi@sns.ias.edu}
\affil{$^2$Institute of Astronomy, Madingley Road, Cambridge CB3 OHA, England;
blackman@ast.cam.ac.uk}
\vskip 0.3cm


\begin{abstract}

We study accretion induced collapse of magnetized white dwarfs as an origin 
of millisecond pulsars. We apply magnetized accretion disk models to the 
pre-collapse accreting magnetic white dwarfs and calculate the white dwarf 
spin evolution. If the pulsar magnetic field results solely from the 
flux-frozen fossil white dwarf field, a typical millisecond pulsar is born 
with a field strength $\sim 10^{11}-10^{12}G$. The uncertainty in the field 
strength is mainly due to the uncertain physical parameters of the 
magnetized accretion disk models. A simple correlation between the pulsar 
spin $\Omega_*$ and the magnetic field $B_*$, $(\Omega_*/10^4s^{-1})\sim 
(B_{*}/10^{11}G)^{-4/5}$, is derived for a typical accretion rate $\sim 
5\times 10^{-8}M_{\sun}/yr$. This correlation remains valid for a wide 
pre-collapse physical conditions unless the white dwarf spin and the binary 
orbit are synchronized prior to accretion induced collapse. We critically 
examine the possibility of spin-orbit synchronization in close binary 
systems. Using idealized homogeneous ellipsoid models, we compute the 
electromagnetic and gravitational wave emission from the millisecond
pulsars and find that electromagnetic dipole emission remains nearly 
constant while millisecond pulsars may spin up rather than spin down
as a result of gravitational wave emission. We also derive the physical 
conditions under which electromagnetic emission from millisecond pulsars 
formed by accretion induced collapse can be a source of cosmological 
gamma-ray bursts. We find that relativistic beaming of gamma-ray emission
and precession of gamma-ray emitting jets are required unless the dipole 
magnetic field strengths are $>10^{15}$G; such strong dipole fields are in 
excess of those allowed from the accretion induced collapse formation 
process except in spin-orbit synchronization. Strong dipole fields 
$>10^{15}$G could in principle be produced in situ. If the millisecond 
pulsars spin up while emitting gravitational wave, the required values for 
the shortest bursts are $>10^{16}$G, pushing the limits of suggested 
neutron star dynamos. 

\end{abstract}


\section{Introduction}

The formation of millisecond pulsars (MSP) is not fully understood.
Explosive supernova events have been considered
as the origin of pulsars (e.g. Mayle \& Wilson 1988, Burrows \& Hayes 1996). 
The high recoil velocities
of the newly formed pulsars (due to the off-center explosions), however,
would likely expel them well away from the Galactic plane 
in contradiction to the observed Galactic MSP distribution. 
An alternative scheme is accretion induced collapse (AIC) of a white
dwarf (e.g. Canal \& Schatzman 1976, Baron et al. 1987, Nomoto \& Kondo
1991, Livio \& Truran 1992).
However, although AIC is the most probable mechanism
for MSP formation (Verbund 1993, van den Heuvel \& Bitzaraki 1995), 
the physics of MSP formation from accreting binary systems 
(Taam \& van den Heuvel 1986, van den Heuvel et al.
1986, Chanmugam \& Brecher 1987, Verbund 1993) and 
the physical conditions  immediately after AIC remain unclear.

AIC may plausibly lead to
pulsars with fast spins in the weakly magnetized white dwarf binary systems
(Chanmugam \& Brecher 1987, van den Heuvel \& Bitzaraki 1995), but
it is not clear how such a mechanism could lead to rapidly rotating and 
very strongly magnetized neutron stars. This is because the strong white dwarf
magnetic field is likely to efficiently brake the  rotation
prior to AIC. Narayan and Popham (1989) investigated the consequence of the 
magnetized accretion and the AIC of the white dwarf 
using the magnetized disk model of Wang (1987). 
They concluded that for an accretion rate ${\dot M}\sim 10^{-7}M_{\sun}/yr$, 
a MSP would form with a typical limiting field strength of 
$\sim 10^{12}G$. Due to recent improvements in the magnetized accretion
disk models (Campbell 1992, Cameron \& Campbell 1993,
Yi 1995, Wang 1995), such qualitative conclusions
can be improved and further quantified.
Some interesting applications of the disk-magnetosphere interaction models
have recently been made to various magnetized accretion systems
(e.g., Yi 1995, Kenyon et al. 1996, Yi \& Kenyon 1996, and references 
therein), where it is found that there could exist a simple spin-magnetic field
correlation even when the disk-star system is not in the equilibrium spin 
state (Yi 1995). If such a simple relation exists in the pulsar systems 
as well, the electromagnetic emission from MSPs is likely to reflect 
such a correlation.

The spin-field parameter space is important because strongly magnetized MSPs 
(Bhattacharya \& van den Heuvel 1991, Taylor \& Stinebring 1986) could be
sources of cosmological phenomena such as cosmological gamma-ray bursts (GRBs) 
(e.g. Usov 1992 and references therein) and gravitational radiation 
(e.g. Backer \& Hellings 1986, Shapiro \& Teukolsky 1983). 
Electromagnetically quiet AIC 
could be as frequent as $\sim 0.1-1 yr^{-1}$ per galaxy
which makes the AIC scenario viable for GRB. 
The quiescent nature of the collapse is crucial for this source of GRBs
because the bursts have not been identified with any precursor
activities (for a review, see, Fishman et al. 1994, Fishman \& Meegan 1995). 
In the Usov (1992) model, the required 
conditions for the pre-collapse white dwarfs are highly extreme compared with
the properties of the known accreting white dwarfs (e.g. Wu \& Wickramasinghe
1993). For a pulsar of
surface dipole magnetic field strength $B_*\sim 10^{15}G$, the moment of inertia
$I_*\sim 10^{45}gcm^2$, the radius $R_*\sim 10^6cm$, and
the spin rate $\Omega_*\sim 10^4s^{-1}$, 
the angular momentum conservation and the 
flux-freezing lead to a pre-collapse white dwarf with the rotational frequency
\beq
\Omega_{wd}\sim \Omega_{*}(I_{*}/I_{wd})\sim 10^{-2}s^{-1}
\eeq
and the magnetic field strength
\beq
B_{wd}\sim B_{*}(R_{*}/R_{wd})^2\sim 10^{9}G
\eeq
for the white dwarf of radius $R_{wd}\sim 10^9cm$ and  moment of inertia
$I_{wd}\sim 10^{51}gcm^2$ (e.g. Frank et al. 1992). 
For the observed cataclysmic variables, such a field 
strength has not been detected (Wu \& Wickramasinghe 1993). 

Usov (1992) suggested a possible origin of  pulsars with 
very strong magnetic fields; a binary system consisting of two dwarfs.
In order for the dwarf binary system to be in mass transfer, the binary 
separation should be close enough to allow the Roche-lobe overflow. 
Based on this observation, Usov (1992) further suggested that the accreting 
magnetized white dwarf's spin is likely to be synchronized with the binary's 
orbital motion through the direct magnetic coupling between the two stars
(e.g. Lamb et al. 1983). 
Such a scenario could indeed give a pre-collapse white dwarf with 
$\Omega_{wd}\sim 10^{-2}$. It has been noted, however, that the mass 
accretion rates required for AIC are as high as 
$\ge 4\times 10^{-8} M_{\sun}/yr$ (Livio \& Truran 1992, Nomoto \& Kondo 1991,
and references therein). It is not clear whether such high accretion rates
could be achieved and sustained for an extended period of time in the 
binary systems where the companion stars are dwarfs. It is therefore crucial to
examine the pre-collapse conditions of the magnetized binary systems.

If the gamma-ray emission is isotropic, 
a cosmological GRB (for a review see e.g. Blaes 1994, 
Fishman et al. 1994, Fishman \& Meegan 1995) source must have the luminosity 
\beq
L_{\gamma}\approx [1.5\times 10^{51} erg/s]
\left(F_{\gamma}\over 5\times 10^{-8} erg/s/cm^2\right)
\left(H_0\over 65km/s/Mpc\right)^{-2}
\left(\chi\over (1-\chi)^{\alpha}\right)^2
\eeq
where $F_{\gamma}$ is the typical detected gamma-ray flux, 
$H_0$ is the Hubble constant, $\chi=1-(1+z)^{-1/2}$, $z\sim 1$ is the
cosmological redshift, and $\alpha\approx 1-2$ is the usual photon index
for gamma-ray spectra (e.g. Blaes 1994, Fishman \& Meegan 1995,
Fishman et al. 1994, Yi 1993 and references therein).
It has been noted, primarily based on energetics and time scales, 
that the MSPs with strong magnetic fields could be interesting 
cosmological sources of electro-magnetic emission (Usov 1992, 1994).
If the pulsars lose their energy through emission of the gravitational 
radiation with luminosity (Shapiro \& Teukolsky 1983), 
\beq
L_{GW}={32 G\over 5c^5}{I_*^2\Omega_*^6}\epsilon^2
\approx [2\times 10^{55} erg/s]\epsilon^2\left(I_*\over 10^{45} g cm^2\right)^2
\left(\Omega_*\over 10^4 s^{-1}\right)^{6}
\eeq
the characteristic energy loss time scale 
\beq
t_{GR}\approx {I_*\Omega_*^2/2\over L_{GR}}\sim [3\times 10^{-3}s]\epsilon^{-2}
\left(I_*\over 10^{45} gcm^2\right)^{-1}\left(\Omega_*\over 
10^{4}s^{-1}\right)^{-4}
\eeq  
could be short enough to account for the observed short durations of the
GRBs (Usov 1992, Blackman et al. 1996).
$I_*$ is the rotational moment of inertia around the 3-axis,
$\Omega_*$ is the rotational frequency of the pulsar, 
$\epsilon^2=e_{12}^2/(2-e_{12}^2)$, and $e_{12}$ is the equatorial 
eccentricity on the plane spanned by the 1-axis and 2-axis perpendicular 
to the rotation 3-axis,
\beq
e_{12}=(1-(R_2/R_1)^2)^{1/2}
\eeq
with the radial extent of the pulsar along the three axes
taken as $R_1,R_2$, and $R_3$ respectively.
For the electro-magnetic radiation relevant for GRB,  
the luminosity (Shapiro \& Teukolsky 1983, Usov 1992,1994, 
Blackman et al. 1996)
\beq
L_{EM}={2\mu_*^2\Omega_*^4\over 3c^3}\approx
[2\times 10^{50} erg/s]\left(R_*\over 10^6cm\right)^6
\left(B_*\over 10^{15}G\right)^2\left(\Omega_*\over 10^4 s^{-1}\right)^4
\eeq
determines the emission time scale only when $e_{12}$ is vanishingly small
or $L_{EM}>L_{GR}$, where $B_*=\mu_*/R_*^3$ is the neutron star's surface 
magnetic field strength and $\mu_*$ is the magnetic moment of the star.
In order for a pulsar to be a cosmological GRB with
short duration and isotropic emission, the pulsar has to be both rapidly 
spinning ($\Omega_*\sim 10^4s^{-1}$) and extremely strongly magnetized
with $B_*\sim 10^{15}G$ (Usov 1992). 

However, we find that MSP with $\ge 10^{15}{\rm G}$ magnetic fields, 
required by the isotropically emitting  MSP cosmological GRB models, 
are incompatible with AIC for observed white dwarf magnetic fields $<10^9$G.
If these pulsars do exist, then their fields 
would have to be generated in-situ by a dynamo. Duncan \& Thompson (1992)
suggest that  dynamos can generate $10^{15}$G  dipole magnetic fields
in millisecond or sub-millisecond pulsars because the
available differential shear and turbulent energy ($\propto \Omega_*^2$)
for conversion to magnetic field is sufficiently large.
However, complications of imposing such a dynamo are addressed
in section 4.  Constraints on the MSPs formed by AIC and the 
correlation between pulsar spin and flux 
frozen fossil field may turn out to be most important.
Even if dynamos could produce $10^{15}$G fields, we will
point out that the shortest bursts require field in excess of
$10^{16}$G if the emission source is electromagnetic dipole radiation and
gravitational radiation from MSPs spins up neutron stars.

Without superstrong fields, cosmological MSP can still be associated 
with GRB by reducing the energy requirements
of the field through relativistic beaming. 
Blackman et al. (1996) proposed that a strongly beamed emission from 
a pulsar with  less extreme physical conditions 
could be the origin of the cosmological GRBs. In this scenario, 
the observed gamma-rays emanate from relativistically beamed jets and the
observed luminosity could be larger than the intrinsic 
luminosity of the source by a factor $\Gamma^2$ due to the relativistic
beaming (Yi 1993) where $\Gamma$ is the relativistic bulk Lorentz factor 
of a gamma-ray emitting jet (cf. Usov 1994). 
For a Lorentz factor of $\Gamma>10^3$, the required 
electro-magnetic power (eq. (1-5)) could be lower by a factor of $>10^6$ and 
hence the required field strength could be $<10^{12}G$ for a MSP.
In this case, however, the characteristic time scale for 
electro-magnetic emission 
\beq
t_{EM}\approx {I_*\Omega_*^2/2\over L_{EM}}\sim [3\times 10^8s]
\left(I_*\over 10^{45} gcm^2\right)
\left(R_*\over 10^6cm\right)^{-6}\left(B_*\over 10^{12}G\right)^{-2}
\left(\Omega_*\over 10^4s^{-1}\right)^{-2}
\eeq
is much longer than the typical GRB durations. Blackman et al. (1996)
pointed out that
precession of the sharply beamed gamma-ray jets is a possible
explanation for the observed short durations of GRBs. In this picture,
the duration of a burst event is determined primarily by the time scale on 
which the jet sweeps by the observer's line of sight. 
An additional slower precession mode, 
naturally expected in the MSP binary systems (e.g. Thorne et al. 1986), 
would account for the absence of repeaters by keeping the beam from
returning to the line of sight.

When the pulsar rotates with a period $P_*=2\pi/\Omega_*$ shorter than 
the critical period (e.g. Chandrasekhar 1969)
\beq
P_{crit}={2\pi\over \Omega_{crit}}\sim [0.7ms]\left(M_*\over 1.4M_{\sun}\right)
^{-1/2} \left(R_*\over 10^6 cm\right)^{3/2},
\eeq
$\epsilon$ could be as high as $\sim 0.1$ as a result of the non-axisymmetric 
gravitational secular instability. 
This result plausibly suggests a 
source of the non-zero quadrupole moment which is the necessary
ingredient for the gravitational radiation emission. The short time scale
$t_{GR}$ could be the GRB duration provided that the
pulsar loses its rotational energy, spins down, and $L_{EM}$ decreases as 
$\propto\Omega_*^4$ (Usov 1992, Blackman et al. 1996). 
It has been, however, unclear how the electro-magnetic
emission evolves during the angular momentum loss by the gravitational
radiation (e.g. Finn \& Shapiro 1990). 
For the pulsars with $\Omega_*>\Omega_{crit}$, the application of the
conventional dipole formula (Ostriker \& Gunn 1969,
Shapiro \& Teukolsky 1983) for the spin-down of the MSPs becomes uncertain 
during the period dominated by the gravitational radiation emission. 
This is because the conventional formula does not take into account the 
change in the moment of inertia (Chandrasekhar 1969). 
In this sense, it remains to be seen how the MSPs 
electromagnetic emission would evolve right after AIC. Such a question
becomes especially interesting for the pulsars with 
$\Omega_*>\Omega_{crit}=2\pi/P_{crit}$, which is often invoked for the
cosmological GRBs (Usov 1992, Duncan \& Thompson 1992) . 

In this paper, we: 
(i) Investigate formation of the MSPs from AIC of accreting white dwarf binary 
systems and determine a spin-magnetic field parameter space for the resulting
pulsars.
(ii) Study electro-magnetic emission from rapidly spinning pulsars 
rotating above the critical frequency.
(iii) Explore the possibility that even though 
AIC produced pulsars do not have strong magnetic fields, the
energy requirement reduction by
relativistic beaming can still allow AIC pulsars to be associated with
cosmological GRBs.
In section 2, we describe the magnetized accretion disk models. Section 3 
gives the results of the AIC and the derived spin-magnetic field correlation.
In section 4, we examine the electro-magnetic emission from the rapidly rotating
pulsars and derive the conditions for the gamma-ray emission from the
MSPs. We conclude in section 5.

\section{Magnetized Accretion Disk Around a  White Dwarf}

We assume that the white dwarf magnetic field is of the dipole type
and is characterized by the stellar magnetic flux $\Phi_{wd}=B_{wd}R_{wd}^2$ 
and magnetic moment $\mu_{wd}=B_{wd}R_{wd}^3$ where 
$R_{wd}$ is the equatorial radius and $B_{wd}$ is the white 
dwarf surface field strength. The vertical component of the magnetic 
field is approximately given by
\beq
B_z(R)=-{\mu_{wd}\over R^3}
\eeq
where we have adopted the usual cylindrical coordinate system $(R,\phi,z)$.
We consider two magnetized accretion disk models in which the accretion disk
is threaded by the white dwarf magnetic field (Ghosh \& Lamb 1979ab,
Wang 1987, Campbell 1992, Yi 1995). We briefly describe the
models with a summary of the relevant formulas. 

The first model
assumes that a single turbulent mechanism is responsible for both  
viscous angular momentum transport and magnetic field diffusive loss 
(Campbell 1992, Yi 1995). 
In this case, using the magnetic 
diffusivity $\eta_t=\alpha v_t H$ where $\alpha$ is the conventional 
Shakura-Sunyaev viscosity parameter (Frank et al. 1992) 
and $H$ is the vertical disk scale height,
the azimuthal component of the field is given by
\beq
B_{\phi}(R)={\gamma\over\alpha}{\Omega_{wd}-\Omega_K\over \Omega_K}B_z(R)
\eeq
where 
$\Omega_{wd}$ is the white dwarf rotational frequency, 
$\Omega_K=(GM_{wd}/R^3)^{1/2}$ is the Keplerian disk rotational frequency
and $\gamma\sim \left| (R/\Omega)(d\Omega/dz)\right|$ 
measures the vertical velocity shear length scale 
between the Keplerian
disk midplane and the corotating stellar magnetosphere (Wang 1987,
Campbell 1992, Yi 1995). 

The larger the ratio $\gamma/\alpha$, the larger the diffusion of
the external stellar field into the disk becomes 
(Ghosh \& Lamb 1979ab, Yi 1995).  
The torque on the star is given as
\beq
N={7\over 6}N_0{1-(8/7)(R_0/R_c)^{3/2}\over 1-(R_0/R_c)^{3/2}}
\eeq
where
$N_0={\dot M}(GM_{wd}R_0)^{1/2}$ and $R_c=(GM_{wd}P_{wd}^2/4\pi^2)^{1/3}$ 
is the corotation radius with $P_{wd}=2\pi/\Omega_{wd}$ (Yi 1995, Wang 1995).
The disk inner radius $R_0$ below which the disk is magnetically disrupted
is determined by
\beq
\left(R_0\over R_c\right)^{7/2}=A\left|1-\left(R_o\over R_c\right)^{3/2}\right|
\eeq
where
$A=2(\gamma/\alpha)B_c^2R_c^3/{\dot M}(GM_{wd}R_c)^{1/2}$ and
$B_c=\mu_{wd}/R_c^3$ (Wang 1987, Yi 1995).
Eq. (2-3) is the combination of the three physically well-defined torque 
components:
In the region between $R_c$ and $R_0$, the disk's Keplerian rotation is
faster than the stellar rotation and the disk exerts the spin-up torque on the
star.
In the region outside $R_c$, the disk rotation is slower than  the stellar
rotation, which results in the spin-down torque. The disk material accreted
by the star adds the specific angular momentum $(GM_{wd}R_0)^{1/2}$.
Using the characteristic parameters, $A$ becomes
\beq
A=[5.79\times 10^7]
\left(\gamma\over\alpha\right)\left({\dot M}\over 10^{17} g/s\right)
^{-1}\left(B_{wd}\over 10^7G\right)^2\left(R_{wd}\over 10^9cm\right)^6
\left(M_{wd}\over M_{\sun}\right)^{-5/3}\left(\Omega_{wd}\over 
1s^{-1}\right) ^{7/3}
\eeq
(Yi 1995).
The torque vanishes when $R_0/R_c=0.9148$. As $R_0/R_c\rightarrow 0$, the
torque asymptotically approaches $N\rightarrow 7N_0/6$.

In the second model (Aly \& Kuijpers 1990, Livio \& Pringle 1992, Wang 1995), 
it is explicitly assumed that the magnetic field pitch (the ratio of
toroidal to vertical field strength  $\equiv |B_{\phi}/B_z|$) 
is limited by the wind-up of the stellar
field lines due to the velocity shear between the Keplerian disk and the
stellar magnetosphere corotating with star (cf. eq. (2-2)). 
This  could be the case if the magnetosphere is nearly force-free 
(i.e. $\nabla \times{\bf B}\propto {\bf B}$) and the
winding-up of the magnetic field lines is limited by reconnection
(Aly \& Kuijpers 1990) within the stellar and disk magnetosphere. 
The details of the magnetosphere are hard to model. But, 
if the reconnection occurs on a time scale of order 
of the vertical shear time scale between the star
and the disk (cf. $\gamma$), then the  pitch of the field will be
limited to a factor of order unity (Aly \& Kuijpers 1990, Livio \& Pringle
1992, Wang 1995). In this case the azimuthal field is 
\beq
B_{\phi}(R)=\gamma_{max}{\Omega_{wd}-\Omega_K\over\Omega_K}B_z(R)
\eeq
for $R\le R_c$ or $\Omega_{wd}\le \Omega_K$ (Wang 1995). 
The parameter $\gamma_{max}$ 
sets the maximum pitch of the field configuration. 
For $R>R_c$ or $\Omega_{wd}>\Omega_K$, we get
\beq
B_{\phi}(R)=\gamma_{max}{\Omega_{wd}-\Omega_K\over\Omega_{wd}}B_z(R).
\eeq

We note that $\gamma_{max}$ of this second model
replaces the ratio 
$\gamma/\alpha$ in the first model for $R>R_c$.
The quantity
$\gamma_{max}$ mainly affects the region outside $R_c$, 
where in the first model the magnetic pitch increases as 
$\sim \Omega_{wd}/\Omega_K\propto R^{3/2}$.
Practically, we will see that 
the difference between the two models is not serious because
the magnetic torque in the outer region rapidly decreases due 
to the steep decrease of the
dipole field $B_z\propto R^{-3}$. 

The inner truncation radius 
$R_0$ is determined by the same equation as for the first model, 
 with $\gamma/\alpha$ 
replaced by $\gamma_{max}$ in the expression for $A$.  We have
\beq
A={2\gamma_{max}B_C^2R_c^3\over {\dot M}(GM_{wd}R_c)^{1/2}}
\eeq
in eqs. (2-4),(2-5). The expression for the torque on the star
becomes (Wang 1995)
\beq
N={7\over 6}N_0{1-(8/7)(R_0/R_c)^{3/2}+(2/21)(R_0/R_c)^3\over
1-(R_0/R_c)^{3/2}}
\eeq
which vanishes ($N=0$) when $R_0/R_c=0.9502$. 

The difference between the
two models is shown in Fig. 1 using the torque behavior as a function
of the disk truncation radius scaled by the corotation radius. 
Note that, as mentioned above, the difference between the two models
is not significant because of the rapid decrease of the dipole field
for $R>R_c$. We therefore expect small differences in
the spin evolution of the white dwarf provided that the binary system
parameters are similar. We take $\alpha=0.1$, $\gamma=1$, and $\gamma_{max}
=1$ and discuss the uncertainties related to these chosen parameters later.

When the mass accretion rates are large and the magnetic fields are 
weak, $R_0$ becomes comparable to the
stellar radius $R_{wd}$ and the accretion disk extends down to the stellar 
surface. In this limit, we use the purely hydrodynamic accretion torque 
${\dot M}(GM_{wd}R_{wd})^{1/2}$ in place of the magnetic accretion
torque (eqs. (2-3),(2-9)). 
Note that in non-magnetized accreting systems, sustained 
accretion is possible even with spin-down when the star spins
near the break-up rate $\sim (GM_{wd}/R_{wd}^3)^{1/2}$ (Popham \& Narayan 
1991, Paczynski 1991). We refer to objects that reach this limit as 
critical rotators (Narayan \& Popham 1989). We, however, do not consider these 
cases in detail because 
their direct collapse to strongly magnetized neutron stars is not likely.

\section{Non-Explosive Accretion Induced Collapse and Neutron Star Spin}

\subsection{Model Parameters for Accretion Induced Collapse}

{\it Initial Mass of White Dwarf:} We assume that the collapse of an O-Ne-Mg 
white dwarf occurs when the central density reaches the electron capture 
threshold for Mg at $\rho_c=3.16\times 10^{9} g/cm^3$ (cf. Nomoto \&
Kondo 1991, Shapiro \& Teukolsky 1983). The critical
Chandrasekhar mass of the white dwarf is taken as $M_{wd}=1.39M_{\odot}$. 
O-Ne-Mg white dwarfs of mass $M_{wd}=1.2-1.37M_{\odot}$ can form from
$8-10M_{\sun}$ progenitor stars (Nomoto \& Hashimoto 1988). We therefore choose 
$M_{wd}=1.25M_{\sun}$ and $M_{wd}=1.35M_{\sun}$ as initial masses of the
white dwarfs (see also Livio \& Truran 1992). 
It turns out that within this range the initial mass 
only significantly affects the time scales to reach collapse for 
a given accretion rate but not the field-spin relation (see below).

{\it Mass accretion Rate:} 
In principle, the nature of the collapse depends on the accretion rate, but
the range of accretion rates, ${\dot M}$, for AIC
has been constrained.
Low mass accretion rates, ${\dot M}<10^{-9}M_{\sun}/yr$, are believed 
to lead to novae or similar eruptions (Paczynski and Zytkow 1978), which would
decrease the mass of the accreting white dwarf (Livio \& Truran 1992). 
Intermediate mass accretion rates, $10^{-9}M_{\sun}/yr<{\dot M}
<4\times 10^{-8}M_{\sun}/yr$, likely lead to
off-center He detonation (Nomoto 1987, Nomoto \& Kondo 1991).
This leaves  high mass accretion rates ${\dot M}>4\times 10^{-8} 
M_{\sun}/yr=2.5\times 10^{18}g/s$ as the most likely window
for AIC (Livio \& Truran 1992). 

For the plausible mass range  
of the O-Ne-Mg white dwarfs
and an accretion rate, ${\dot M}\sim 10^{-8}M_{\sun}
/yr$  of  observed white dwarf binary systems, the accretion rate
has to be sustained  for $\ge 10^6yr$. 
Such accretion rates are generally expected
from a mass-losing lower giant branch star of mass $\sim 1M_{\sun}$ 
that climbs up the giant branch (Webbink et al. 1983). It is 
questionable whether such a high accretion rate could be achieved for
a dwarf binary system where the secondary star in the binary has a mass
$\ll 1M_{\sun}$ (cf. Usov 1992). For most of the observed white dwarf binary 
systems (cataclysmic variables), the mass accretion rates generally appear to 
be much lower than the typical Eddington rate 
${\dot M}\le 1.4\times 10^{19}g/s$ (e.g. Frank et al. 1992). 
This may imply a narrow range of allowed mass accretion rates
for pre-AIC binary systems. Although very high mass accretion
rates may also lead to collapse, they would be very rare and the duration
of such activity in the binary systems is most likely to be  short
(e.g. Smak 1984).
We therefore mainly consider mass accretion rates near ${\dot M}=5\times 
10^{-8}M_{\sun}/yr$. Although we consider a larger rate ($10^{-7}M_{\sun}/yr$)
and a smaller ($2\times 10^{-8}M_{\sun}/yr$) rate for comparison, the differences 
do not affect our conclusions.

{\it Rotating White Dwarfs:} 
Our calculations require 
a rotational law, radial density profile,  and moment of inertia.
Most importantly, we need a mass-radius relation as the mass 
increases through accretion and the star spins up. We specifically adopt
the results in Hachisu (1986) which are relevant for uniformly rotating
white dwarfs (e.g. Narayan \& Popham 1989). 
Here the white dwarf as a whole is approximated as a rigidly 
rotating solid body. When the white dwarf is rapidly spinning, 
the eccentricity of the star with respect to the rotation axis becomes 
significant. The results of Hachisu (1986) can be interpreted essentially
as a family of solutions corresponding to a combination
of parameters, $M_{wd}$, angular momentum $J_{wd}$, $R_{wd}$, and
$R_3/R_1$ (cf. eq. (1-6)) for a given central density of the axisymmetric star.

{\it Spin Evolution:}
When the white dwarf mass increases significantly through accretion and
the radius of the rapidly rotating star varies in response,
the spin of the star is affected both by the accretion disk torque
and by the change in the moment of inertia.  
Angular momentum conservation gives
\beq
{dJ_{wd}\over dt}=\Omega_{wd}{dI_{wd}\over dt}+I_{wd}{d\Omega_{wd}\over dt}=N
\eeq
where $J_{wd}=I_{wd}\Omega_{wd}=\beta M_{wd}R_{wd}^2\Omega_{wd}$, 
the moment of inertia $I_{wd}=\beta M_{wd} R_{wd}^2$ and
the constant $\beta=0.08$ depends on the details of the white dwarf 
structure including the rotational flattening and the radial density 
profile (Hachisu 1986). Using this expression, we get
\beq
{d\Omega_{wd}\over dt}=\Omega_{wd}\left[{N\over\beta M_{wd}R_{wd}^2\Omega_{wd}}
-{1\over M_{wd}} {dM_{wd}\over dt}-2{1\over R_{wd}}{dR_{wd}\over dt}\right]
\eeq
or
\beq
{d\Omega_{wd}\over dt}=\Omega_{wd}\left[{N\over J_{wd}}-{1\over M_{wd}}
{dM_{wd}\over dt}- 2{1\over R_{wd}}{dR_{wd}\over dt}\right]
\eeq
We further assume that the magnetic flux threading the stellar surface is
conserved, i.e.
\beq
B_{wd}=B_{wd,i}\left(R_{wd,i}\over R_{wd}\right)^2
\eeq
where the subscript $i$ denotes the initial epoch at which the accretion and
spin evolution begin. Finally, we assume that all white dwarfs 
spin sufficiently
slowly at the beginning of their evolution, so that, due to the short
spin-up (-down) time scales, the details of the initial spin rates are
not necessary (e.g. Yi 1995).
Once the constant mass accretion rate is chosen and the initial field
strength is assigned, the mass increase and torque are calculated while the 
radius and the magnetic field of the white dwarf are updated accordingly.

There are three possible cases for collapsing white dwarfs (e.g. Narayan
and Popham 1989):

(i) When the star rotates faster than the critical rate, roughly equal
to the equatorial Keplerian rotation $\sim (GM_{wd}/R_{wd}^3)^{1/2}$,
the star can lose its mass due to the centrifugal force (i.e. equatorial
mass shedding) and the rotational support prevents any further collapse.
If the mass is continuously added to the star with the critical rotation, 
it is unclear how the star would respond. 
In the non-magnetic, steady state case, the purely hydrodynamic accretion can
proceed with continuous {\it spin-down} as shown by Popham and Narayan (1991) 
and Paczynski (1991). 
In the magnetized accretion case, the accretion occurs mainly
along the polar accretion column (Frank et al. 1992). 
This polar accretion may continue
despite the loss of matter near the equatorial plane. Given this complex
situation, we simply assume that the equatorial mass loss and the
polar inflow balance, so that the net accretion essentially ceases
 when the star reaches
the critical rotation. 

For comparison with  other possible evolutions, 
we calculate the hypothetical MSP spin and magnetic field after
a flux freezing and angular momentum conserving collapse of the white dwarf
at the critical rotation. Practically, these objects
correspond to those with weak magnetic fields, which are about $\sim 80\%$
of the known cataclysmic variables (Morris et al. 1987), 
and hence they are unlikely
to be a source of  intense electromagnetic emission. Such MSPs, if
they indeed form from AIC, could be a significant source of the gravitational
radiation.

(ii) Consider now the case for which the white dwarf reaches the collapse 
density and the angular momentum of the star is below the critical rotation 
limit but above  
$\sim 6\times 10^{48} gcm^2/s$. Here the collapse of the star
is likely to be halted after shrinking to a size for which the centrifugal
forces prevent further collapse. These objects are often referred as
fizzlers (Shapiro and Lightman 1976, Tohline 1984). 
The critical angular momentum for the fizzlers is taken approximately as
$J_*=I_*\Omega_*\approx 6.3\times10^{48}g cm^2/s$ for
$P_*=2\pi/\Omega_*\approx 10^{-3}s$. Fizzlers with fast spins are likely to
be unstable to non-axisymmetric perturbations (Lai \& Shapiro 1995). The resulting 
triaxial
configurations could lose angular momentum through gravitational radiation 
which could push the object to below the critical angular momentum limit.
The collapse would then resume leading to the formation of a pulsar.
Below, we treat the collapse of fizzlers as direct collapses
by neglecting possible angular momentum loss at the fizzler stage
(cf. Durisen et al. 1986, Williams \& Tohlin 1988, Houser et al. 1994).
In this case, we also expect relatively small magnetic fields for most of
the fizzlers and these cases have only limited relevance for
strong electromagnetic emission and any associated cosmological GRBs.

(iii) Finally, when the angular momentum is below the fizzler limit and the 
central density exceeds the critical value $\rho_c\sim 10^{9.5} g/cm^3$
(Nomoto \& Kondo 1991, Shapiro \& Teukolsky 1983),
the stars could collapse directly to neutron stars. In this case, the
field strength is expected to be large and the rotation is relatively slow.
We assess how likely it is to get rapidly spinning strongly magnetized neutron 
stars in this case.  We calculate the final spin and magnetic field
assuming conservation of angular momentum and magnetic flux. 
The magnetic flux of the collapsed object is then given by 
$B_*=(R_{wd}/R_*)^2B_{wd}$ and the rotational velocity $\Omega_*=(I_{wd}/I_*)
\Omega_{wd}$ where we take $R_*=10^6cm$ and $I_*=10^{45}gcm^2$. 

\subsection{Correlation between Spin and Magnetic Field}

Fig. 2 shows the main results of the spin evolution. In all panels,
the dashed lines correspond to the cases where the white dwarfs reach
the critical rotation before collapse. The field strengths and spins
for the critical rotators shown in Fig. 2 are hypothetical values (strictly
for comparison) based on the assumption that the stars begin to collapse
at the critical rotation. These objects are 
characterized by fast spins and weak magnetic fields reflecting
the inefficient magnetic braking of the disk-star interaction.
The dotted lines correspond to the fizzlers which are assumed to
collapse down to the neutron stars after reaching the critical central
density. Since the fizzlers arrive at the collapse line with
more angular momentum than the objects which collapse directly
(solid line), the neutron stars formed from fizzlers would be observed
as those with somewhat higher spins and weaker magnetic fields. In our
models, the resulting magnetic fields are not strong enough for the
fizzlers to be strong emitters of electromagnetic radiation but they
could be strong sources of gravitational radiation.  

In all panels, thin lines are the results of the pitch-limited magnetized
disk model (eq. (2-9)) and the thick lines are for the model with the 
diffusive magnetic field loss (eq. (2-3)).
In the upper panels, the initial mass of the white dwarf is 
$M_{wd}=1.25M_{\sun}$.
>From (a) to (c), the mass accretion rate increases from $2\times 10^{-8}
M_{\sun}/yr$ to $10^{-7} M_{\sun}/yr$. The highest accretion
rate results in the fastest spin. We note that the highest mass accretion 
rate used in (c) has not been seen in typical white dwarf binary systems 
such as the cataclysmic variables (Frank et al. 1992) except during
brief outbursts (Smak 1984). Although some transient phases
of high mass accretion rates cannot be ruled out, such high accretion
rates are less likely to be sustained for an extended period of time. 

In the lower panels (d)-(f), the initial mass of the white dwarf
is increased to $M_{wd}=1.35M_{\sun}$. The differences between the upper 
and lower panels, due to the difference in the white dwarf mass, are 
significant only for the critical rotators.
We observe that the typical high spin pulsars 
$\Omega_*\sim 10^{4} s^{-1}$ would be born with $B_*\sim 10^{11}G$.
This result is qualitatively consistent with the result of Narayan \&
Popham (1989). 
The derived correlation between spin and magnetic field strength 
is one of  the main quantifiable characteristics of the MSPs formed from AIC.
Different choices of the magnetic field model and the initial white dwarf
mass only slightly affect the results. 

Since there is little difference between results in the two different 
white dwarf initial mass cases, we can write down a simple spin-magnetic 
field correlation for a given mass accretion rate. The two distinctive 
magnetic field models we discussed in section 2 also do not result in serious
numerical differences. For the diffusive loss case (thick lines) and for
the mass accretion rate ${\dot M}=5\times 10^{-8}M_{\sun}/yr$,
the results in Fig. 2 show that the spin-field correlation gradually varies
from $\Omega_*\propto B_*^{-0.73}$ near the fizzler regime to
$\Omega_*\propto B_*^{-6/7}$ toward the slow spin regime. 
The latter correlation is precisely what is expected in the spin equilibrium
situation (cf. eqs. (2-3),(2-5)) before AIC; the equilibrium relation is 
self-similar because both the field and the angular velocity vary inversely 
with $R^2$.

The results indicate that the spin equilibrium is
not achieved for most of the rapidly spinning white dwarfs.  This is not 
surprising given the fact that the mass and moment 
of inertia are constantly changing
during the pre-collapse accretion phase. For MSP (or sub-MSPs, i.e. near
(i.e. near $\Omega_*\sim 10^4 s^{-1}$),  and
moderately strong magnetic fields, the spin-field correlation is
best described by 
\beq
\left(\Omega_*\over 10^4 s^{-1}\right)\approx 1.8\left(B_*\over 10^{11}G\right)
^{-4/5}
\eeq
where we again note that the two initial masses have little effects on the
cases of the fizzlers and direct collapses. The initial white dwarf mass
only affects the early spin evolution to critical rotation before collapse. 

For the pitch-limited torque model (thin lines) and
for the same parameters as above we get
\beq
\left(\Omega_*\over 10^4s^{-1}\right)\approx 6.3\left(B_*\over 10^{11}G\right)
^{-4/5}.
\eeq
This simple spin-field correlation is one of the main results of the paper.
An analogous spin-field correlation has also been predicted in the
magnetized accretion model for the protostellar systems (Yi 1995, Kenyon et al.
1996).

There are two reasons for the small differences between the two torque 
model predictions.  The first is that in model two, 
the assumed maximum pitch $\gamma_{max}$ effectively reduces the 
spin-down torque contributed by the region outside the corotation radius. 
This effect is less important as the dipole field strength rapidly 
decreases outside the corotation radius and the torque contribution 
from the outside region
is relatively small. The second reason is due to the scaling in the effective 
measure of the magnetic field strength. 
For the diffusive loss model, we have 
adopted $\alpha=0.1$ and $\gamma=1$ which 
combines to give $\gamma/\alpha=10$. As we have mentioned
earlier, $\gamma_{max}$ essentially has the same physical effect as
the combination $\gamma/\alpha$. Therefore, for our choice of
$\gamma/\alpha=10$, the effective field strength in the diffusive loss
model is roughly three times larger than that of
the magnetic pitch-limited model.
The results in Fig. 2 reflect this effect and show smaller spin
rates for the same stellar field strength and the mass accretion rate. 

Although the spin equilibrium ($N=0$) condition is not always reached during 
the pre-AIC spin evolution shown in Fig. 2,
the equilibrium spin provides a useful scaling for the derived spin-field
correlation and especially for the dependence on the model parameters 
$\gamma/\alpha$ or $\gamma_{max}$.
Assuming that the final pre-collapse white dwarf spin period is
the equilibrium spin period, using eq. (2-5), we can show that
the pre-collapse white dwarf parameters have to satisfy
\beq
\left(\gamma\over\alpha\right)\left(B_{wd}\over 10^7G\right)^2
\left(R_{wd}\over 10^9cm\right)^6\left(M_{wd}\over M_{\sun}\right)^{-5/3}
\left(P_{wd}\over 100s\right)^{-7/3}\left({\dot M}\over 10^{18}g/s\right)^{-1}
\approx 6.5\times 10^{-4}
\eeq
which shows the equilibrium spin-field correlation 
$P_{wd}\propto B_{wd}^{6/7}$ as noted earlier.
For the representative parameters $R_{wd}\sim 10^{9}cm$,
$M_{wd}\sim 1.3M_{\sun}$, $\Omega_{wd}\sim 10^{-2}s^{-1}$ relevant for
AIC, 
\beq
\sqrt{\gamma\over\alpha}\left(B_{wd}\over 10^6G\right)\approx 3\left({\dot M}
\over 10^{18}g/s\right)^{1/2}
\eeq
which is consistent with the derived result, eq. (3-5), within a factor of 
order unity. That is, assuming flux-freezing and angular momentum
conservation for AIC, the derived spin-field correlation eq. (3-5) can
be used in eq. (3-7), which results in a relation very close to eq. (3-8).
Taking into account the anticipated uncertainty in $\gamma/\alpha$, for
instance, by about an order of magnitude, our derived field strengths would be 
uncertain at most by a factor of $\sim 3$. 
Typically, for the cataclysmic variables $\alpha\sim
0.1$ is often quoted (Frank et al. 1992). 
$\gamma$ could in principle range from $\sim 1$ to 
\beq
\gamma\sim {R\over H}\sim 10\left(\alpha\over 0.1\right)^{1/10}
\left({\dot M}\over 10^{18}g/s\right)^{-3/10}\left(M_{wd}\over M_{\sun}\right)
^{3/8}\left(R\over 10^{10}cm\right)^{-1/8}
\eeq 
where $H$ is the vertical thickness of the accretion disk (e.g. Frank et al.
1992). This indicates $\gamma\sim O(1)$ for high mass accretion rates
${\dot M}>3\times 10^{18}g/s$ appropriate for AIC.
In the case of the pitch-limited case, from eq. (2-8), 
we arrive at a result similar to
eq. (3-8) with $\gamma_{max}$ in place of $\gamma/\alpha$, i.e.
\beq
\sqrt{\gamma_{max}}\left(B_{wd}\over 10^6G\right)\approx
3\left({\dot M}\over 10^{18}g/s\right)^{1/2}
\eeq
which is also consistent with eq. (3-6) within a factor of order unity.
We note that $\gamma_{max}$ is expected to remain close to unity
as long as the reconnection time scale is close to the time scale
of the vertical velocity shear (Aly \& Kuijpers 1990).
Eqs. (3-8) and (3-10) are consistent with the dependence of the spin-field
correlation on $\gamma/\alpha$ or $\gamma_{max}$ seen in eqs. (3-5) and 
(3-6). We conclude that the numerical difference between the two spin-field 
correlations is largely due to the different effective strength of the
magnetic field determined by $\gamma/\alpha$ or $\gamma_{max}$.
It is gratifying that the uncertainty in $\gamma$ and $\gamma_{max}$ is
likely to be at most at the level of order unity.

\subsection{Star-Star Direct Magnetic Coupling}

We now briefly examine the possibility of magnetic coupling between the
accreting white dwarf and the mass-losing secondary.
Such  magnetic coupling  between pre-collapse, close binary stars
has to be considered because it may lead to a synchronization
of  the white dwarf's
spin with the orbital motion (Lamb et al. 1983, Lamb
and Melia 1987). As pointed out by Usov (1992), the short spin period
of the pre-collapse white dwarf could be due to this type of magnetic
coupling. A critical assessment in the context of AIC is necessary.

The magnetohydrodynamic torque between the white dwarf and the orbiting 
secondary is estimated to be (Lamb et al. 1983)
\beq
N_{bin}\sim \phi\gamma_b D R_2^2 (\mu_{wd}/D^3)^2
\eeq
where $\phi\le 1$ is the fractional area of the secondary star threaded by the
magnetic flux, $\gamma_{b}$ is the pitch of the magnetic fields connecting
the two stars, $D$ is the binary separation, and $R_2$ is the 
secondary star's radius.
Assuming $\gamma_b\sim 1$ (Low 1982, Aly 1984), we can estimate
the importance of the magnetic torque due to the secondary. Once
the system is synchronized, we expect $\gamma_b=0$. 
If the secondary star is magnetically active, $\phi\sim 1$ is expected 
and we can use this as an upper limit. We estimate
\beq
N_{bin}\sim [3\times 10^{33} G^2cm^3](\phi\gamma_b)
\left(D\over 5\times 10^{10}cm \right)^{-5}
\left(R_2\over 3\times 10^{10}cm\right)^2
\left(\mu_{wd}\over 10^{33}Gcm^3\right)^2
\eeq
where $\mu_{wd}=B_{wd}R_{wd}^3\sim 10^{33} Gcm^3$ is the magnetic moment of 
the white dwarf
with $R_{wd}\sim 10^9cm$ and $B_{wd}\sim 10^6G$. We note however that
$\mu_{wd}$ could be as high as $10^{36}Gcm^3$ for very strongly magnetized
white dwarfs with $B_{wd}\sim 10^9G$ as suggested by Usov (1992). Although such
white dwarf magnetic fields $\sim 10^9G$ have not been directly observed,
the possibility that a small fraction of magnetized white dwarfs
may have such strong fields cannot be ruled out.
We compare $N_{bin}$ with the typical magnetic torque due to the accretion disk
\beq
N\sim [1\times 10^{36} G^2cm^3]
\left({\dot M}\over 10^{18}g/s\right)\left(M_*\over M_{\sun}
\right)^{1/2}\left(R_0\over 3\times 10^9cm\right)^{1/2}
\eeq
(cf. eqs. (2-3),(2-9)). $N_{bin}$ becomes comparable to $N$ only when
$D\ll 10^{11} cm$ or {\bf $B_{wd}\gg 10^6G$.}

In order for the accretion through the Roche lobe flow to 
occur, the binary separation $D$ has to satisfy
\beq
D<2R_2\left(q\over 1+q\right)^{1/3}
\eeq
where $M_2$ is the mass of the secondary star and the binary mass ratio 
$q=M_2/M_{wd}$ (Frank et al. 1992, Paczynski 1971). 
For $N_{bin}\gg N$ or $D\ll 10^{11}cm$, 
$M_2\ll 0.7[q/(1+q)]^{-0.36}$. We have assumed
the dwarf mass-radius relation for the secondary together with the binary 
orbit formula originally due to Paczynski (e.g. Paczynski 1971). We 
do not consider massive companions due to their short evolution time scales. 
The synchronization of the white 
dwarf spin with the binary orbit could occur on a time scale
$$
t_{sync}\sim {I_{wd}\Omega_{wd}/N_{bin}}
\qquad\qquad\qquad\qquad\qquad\qquad\qquad\qquad\qquad\qquad
$$
\beq
\sim [3\times 10^{7}yr]
(\phi\gamma_b)^{-1}
\left(I_{wd}\over 3\times 10^{50}gcm^2\right)
\left(\Omega_{wd}\over 10^{-2}s^{-1}\right)
\left(\mu_{wd}\over 10^{33}Gcm^3\right)^{-2}
\left(D\over 5\times 10^{10}cm\right)^5
\left(R_2\over 3\times 10^{10}cm\right)^{-2}.
\eeq
The white dwarf spin would be synchronized with the binary orbital frequency
\beq
\Omega_{bin}=2\pi/P_{bin}\sim [10^{-3}s^{-1}]\left[(M_{wd}+M_2)\over 1.5M_{\sun}
\right]^{1/2}\left(D\over 5\times 10^{10}cm\right)^{-3/2}
\eeq
which is interesting only when the secondary star is a rather extreme
dwarf star.  Otherwise, the accretion and collapse would have taken place
by the time the synchronization could have taken place.
Assuming that the accreted mass is typically $\sim 0.1M_{\sun}$ 
for collapse, which we take as the minimum mass for the secondary, 
the typical size of such a dwarf star would be 
$R_2\sim 10^{10}cm$ using the dwarf 
mass-radius relation. That is, unless the binary stars are in physical contact
(i.e. a contact binary),
the torque from the accretion disk  dominates 
in white dwarf binary systems with less extreme conditions.  
We conclude that the accretion disk torque is dominant except
when the binary separation $D<10^{10}$cm for ${\dot M}>10^{18}$ g/s
provided that white dwarfs have magnetic fields not much stronger than
the typical field strength $\sim$ a few $\times 10^6G$.

If the binary stars get close enough to have a short
synchronization time scale, $t_{sync}$,
the orbital evolution due to the gravitational 
radiation emission is not negligible. The orbital evolution time scale
due to the gravitational radiation emission is
\beq
t_{orb,GR}=\left|{1\over P_{orb}}{dP_{orb}\over dt}\right|^{-1}
\approx [2\times 10^5yr]\left(\Omega_{bin}\over 10^{-2}s^{-1}\right)^{-8/3}
\left(M_{wd}\over M_{\sun}\right)^{-8/3}{(1+q)^{1/3}\over q}
\eeq
which is shorter than the synchronization time scale unless the binary
separation $D$ is much shorter than $\sim 5\times 10^{10}cm$. 
For $B_{wd}\sim 10^6$G, the synchronization due to the direct star-star magnetic
coupling is most likely to operate in systems with ${\dot M}\ll 10^{18}g/s$, 
which is not relevant for AIC. 
We note however that the condition $t_{sync}<t_{orb, GR}$ could be easily 
met if white dwarfs are very strongly magnetized with $B_{wd}\gg 10^6$G
(eqs. 3-15,3-17). In such a case, $N_{bin}>N$ (eqs. 3-12, 3-13)
and the spin-orbit synchronization is achieved.

Observationally, the spin-orbit synchronization occurs in AM Her systems, which
are a subclass of the cataclysmic variables with relatively strong magnetic
fields $\sim 10^7G$. The typical white dwarf spin periods $\sim 10^4s$ would 
certainly be too long to produce $P_*\sim 10^{-1}s$ in the event of AIC. 
The spin periods shorter than $\sim 10^4s$ are mostly observed in the DQ Her
systems which are relatively weakly magnetized with $B_{wd}<10^7G$, which
indicates that the short spin periods are more likely in systems with accretion
disks and weak fields (Warner \& Livio 1987, Warner \& Wickramasinghe 1991). 
This observation is consistent with our estimates on the
relative importance of the accretion torque, eqs. (3-12),(3-13).
As noted by Chanmugam \& Brecher (1987), the observed
cataclysmic variables have no trouble being candidates for the MSPs through
AIC. In this sense, the role of the star-star synchronization is not essential
in producing MSPs. The known number of cataclysmic variables $\sim 10^5$ in
our Galaxy (de Cool \& van Paradijs 1987) indicates that only a small fraction 
of them are required to undergo AIC to produce $\sim 10^2$ low mass X-ray 
binaries with neutron stars (Chanmugam \& Brecher 1987). 

\section{Electromagnetic Emission from Rapidly Rotating Pulsars}

\subsection{Electromagnetic Emission and Relativistic Beaming}

Our derived correlation between the spin and the magnetic field
strength indicates that the electromagnetic dipole
emission from MSP immediately after AIC has a very 
simple scaling. The dipole emission formula eq. (1-5),
\beq
L_{EM}\propto B_*^2\Omega_*^4
\eeq
leads to a power-spin correlation
\beq
L_{EM}\propto B_*^2\Omega_*^4\propto B_*^{-6/5}\propto\Omega_*^{3/2}
\eeq
where we have used the spin-field correlation eqs. (3-5),(3-6).
Taking into account the numerical factors in eqs. (3-5) and
(3-6), we write the single scaling for the spin-field correlation
\beq
\left(\Omega_*\over 10^4s^{-1}\right)\approx 
\left(B_{eff}\over 7\times 10^{11}G\right)^{-4/5}
\eeq
where $B_{eff}=\delta B_*$ with $\delta=\sqrt{\gamma/\alpha}$ or 
$\delta=\sqrt{\gamma_{max}}$ depending on the model. 
The electromagnetic power from the pulsars is then given by
\beq
L_{EM}\sim [10^{44} erg/s]\delta\left(R_*\over 10^6cm\right)^6
\left(\Omega_*\over 10^4s^{-1}\right)^{3/2}.
\eeq
This is  much smaller than that required in eq. (1-3) 
indicating that GRB emission would require a high degree of beaming
or anisotropy (Blackman et al. 1996); otherwise the burst event would not 
be visible at cosmological distances. 
Assuming a nominal
gamma-ray emission efficiency $\xi=0.1$, the observed gamma-ray luminosity
$L_{\gamma}>10^{51} erg/s$ would require a degree of the relativistic
beaming characterized by the bulk Lorentz factor $\Gamma$,
\beq
\Gamma=(L_{\gamma}/\xi L_{EM})^{1/2}>10^4 \delta^{1/2}(\xi/0.1)^{-1/2}
\eeq
for a typical MSP with $\Omega_*\sim 10^4 s^{-1}$ and $R_*\sim 10^6cm$.
This is the first necessary ingredient of the present cosmological GRB
scenario.

The low intrinsic
power also presents another problem as pointed out in Blackman et al. (1996).
Using the expected electromagnetic power, eq. (4-4),
the characteristic duration of the electromagnetic emission is (cf. eq. (1-8))
\beq
t_{EM}={I_*\Omega_*^2/2\over L_{EM}}
\sim [5\times 10^{8}s]\left(R_*\over 10^6cm\right)^{-6}
\left(I_*\over 10^{45} gcm^2\right)
\left(\Omega_*\over 10^4s^{-1}\right)^{1/2}
\eeq
if the pulsar's rotational energy is lost only through the dipole radiation. 
When the pulsar's rotation is fast enough to trigger the gravitational
secular instability (Chandrasekhar 1970ab), the pulsar's quadrupole moment can
be significant and the pulsar's angular momentum will be lost on a time scale
$\sim t_{GR}$ (eq. (1-5)). If $I_*$ remains constant during the emission of the 
gravitational radiation, the characteristic time scale of spin-down 
could be $\sim t_{GR}$ on which $L_{EM}\propto \Omega_*^4$ would also
rapidly decrease. This time scale, eq. (1-5), which depends sensitively
on $\epsilon\sim 0.1$,
is short enough for the short duration cosmological GRBs (Usov 1992)
if rapid spin-down occurs as a consequence of gravitational radiation
emission.
The change of inertia as a function of angular momentum is however far more
complex than the simple case of $I_*=constant$ (e.g. Chanrasekhar 1969, Finn \&
Shapiro 1990) and the spin evolution during the gravitationally unstable
phase may cause spin-up rather than spin-down. In this sense, it is not
clear how $L_{EM}$ would evolve and what the characteristic time scale
for the evolution of $L_{EM}$ is. We therefore must ask:  what is the
characteristic time scale for variation of the electromagnetic
power during the evolution driven by gravitational radiation emission? 

\subsection{Magnetized Neutron Stars as Homogeneous Ellipsoids}

We address this question by approximating the pulsars as
magnetized homogeneous ellipsoids (Chandrasekhar 1969).
When a homogeneous ellipsoid rotates with a period shorter than the critical 
period eq. (1-9), the axisymmetric Maclaurin spheroids bifurcate into
two branches (e.g. Chandrasekhar 1969, Shapiro \& Teukolsky 1983). 
In addition to the continuous sequence of the Maclaurin
spheroids,  non-axisymmetric ellipsoids also exist as an equilibrium
configuration.  
Maclaurin spheroids with faster than critical spins are 
unstable to non-axisymmetric secular instability and 
triaxial Jacobi ellipsoids
have often been used to model the equilibrium configuration resulting from
this instability. The non-zero quadrupole moment of a
non-axisymmetric star is the origin of the gravitational  
radiation.
As shown by Chandrasekhar (1970ab), the equilibrium Jacobi
ellipsoids can spin up even though angular momentum is radiated away.
This will strongly affect the
electromagnetic emission because the dipole emission formula 
involves the magnetic moment and the spin rate, not the total angular momentum.
When the pulsar rotates at a rate below the critical rate, the non-axisymmetry
is irrelevant and the electromagnetic emission gives a long time scale
(eq. (4-6)). But, if the pulsar rotates faster than critical, 
the pulsar's overall electromagnetic emission time scale
has to be separately considered.

A large non-axisymmetry can also result from  deformation by
strong enough magnetic fields  (Ostriker \& Gunn 1969, Usov 1992). This may 
affect the properties of pulsars' equilibrium configuration and may change
the bifurcation point at which the Jacobi ellipsoids branch out 
and hence the eventual gravitational radiation emission (Tsvetkov 1983, 
Bonazzola \& Gourgoulhon 1996). When the magnetic fields are very strong,
the equatorial eccentricity caused by the asymmetric magnetic tension
is estimated as
\beq
\epsilon\sim {R_*^4\over GM_*^2}\left(3B_p^2-B_{\phi}^2\right)
\sim 10^{-4}{B_p^2-(1/3)B_{\phi}^2\over (3\times 10^{15} G)^2}
\eeq
where $B_p$ is the poloidal field component and $B_{\phi}$ is the
toroidal field component. Therefore, for the fields we have derived
$\sim 10^{11-12}G$, such deformation effects should be  negligible. 
The pulsar is treated as a uniformly rotating homogeneous ellipsoid. 
We consider the dynamical
evolution of the ellipsoid and assume that magnetic field responds passively
to the change in pulsar configuration.

Following Chandrasekhar and Esposito (1970), the energy loss rate of
the pulsar due to gravitational radiation is given by
\beq
{dE_*\over dt}=-{G\over 45c^5}\left<{d^3D_{ij}\over dt^3}\right>
\left<{d^3D_{ij}\over dt^3}\right>
\eeq
and the angular momentum loss is given by
\beq
{dJ_*\over dt}={4G\over 5c^5}\left<{d^3I_{ik}\over dt^3}{d^2I_{ik}\over dt^2}
\right>
\eeq
where $E_*$ is the total energy of the pulsar, $J_*$ is the angular momentum,
$D_{ij}=3I_{ij}-\delta_{ij}I_{kk}$, and $I_{ij}$ is the moment of
inertia tensor
given by the standard formula (Chandrasekhar 1969)
\beq
I_{ij}=\int_v\rho_* x_i x_jd^3x
\eeq
where $\rho_*$ is the uniform stellar density and ${\bf x}$ is the
displacement vector from the center.
We have adopted the standard tensor notation with $i,j,k$
denoting coordinate indices. The $\left<\right>$ denotes averaging.

For a quasi-steady pulsar in the frame rotating with  angular velocity
$\Omega_*$, eqs. (4-8),(4-9) become
\beq
{dE_*\over dt}=-{32G\Omega_*^6\over 5c^5}\left(I_{11}-I_{22}\right)^2
\eeq
\beq
{dJ_*\over dt}=-{32G\Omega_*^5\over 5c^5}\left(I_{11}-I_{22}\right)^2
\eeq
which recovers the familiar equation
\beq
{1\over \Omega_*}{dE_*\over dt}={dJ_{*}\over dt}.
\eeq
The triaxial Jacobi spheroids have axes
$R_1$, $R_2$, and $R_3$,  related by the
constraint equations
\beq
\left(R_1 R_2\over R_3\right)^2=\left(A_3\over A_{12}\right)
\eeq
and
\beq
\Omega_*^2=2B_{12}
\eeq
where
\beq
A_{12}=R_1R_2R_3\int_0^{\infty}{dr\over (R_1^2+r)^{3/2}(R_1^2+r)
^{3/2}(R_2^2+r)^{1/2}}
\eeq
\beq
A_3=R_1R_2R_3\int_0^{\infty}{dr\over (R_1^2+r)^{1/2}(R_2^2+r)^{1/2}
(R_3^2+r)^{3/2}}
\eeq
\beq
B_{12}=R_1R_2R_3\int_0^{\infty}{rdr\over (R_1^2+r)^{3/2}(R_2^2+r)^{3/2}
(R_3^2+r)^{1/2}}
\eeq
and we have taken the 3-axis as the rotation axis.
Another constraint  comes from the assumed homogeneity and
incompressibility, i.e.
\beq
R_1R_2R_3={\bar R}^3=constant
\eeq
We define the axis ratios $l_2=R_2/R_1$ and $l_3=R_3/R_1$ from which we get
$R_1={\bar R}(l_2l_3)^{-2/3}$.
The coupled differential equations for the evolution become
\beq
{d\l_2\over d\tau}={C_3C_5\over C_1C_5-C_2C_4}
\eeq
\beq
{dl_3\over d\tau}={C_3C_4\over C_2C_4-C_1C_5}
\eeq
where $\tau=t/t_G$ is the dimensionless time in units of the characteristic
gravitational time (Chandrasekhar 1970ab)
\beq
t_G={25\over 18}\left({\bar R}\over R_s\right)^3{{\bar R}\over c}
\approx [6.6\times 10^{-4}s]\left(M_*\over 1.4M_{\sun}\right)^3\left({\bar R}
\over 10^6cm\right)^4
\eeq
where $R_s=2GM_*/c^2$ is the Schwarzschild radius. The relevant quantities
used in eqs. (4-20),(4-21) are
\beq
C_1=l_3\left[{1\over 3}(11l_2^2-1)X_1-3(1+l_2^2)X_2\right]
\eeq
\beq
C_2=-l_2(1+l_2^2)\left({1\over 3}X_1+X_3\right)
\eeq
\beq
C_3=-(l_2l_3)^{1/2}(1-l_2^2)^2X_1^3
\eeq
\beq
C_4=l_2\left(3X_2-{l_3^2\over l_2^2}X_3\right)
\eeq
\beq
C_5=-l_3\left(3X_5-{l_2^2\over l_3^2}X_3\right)
\eeq
with
\beq
X_1=l_2l_3\int_0^{\infty}{rdr\over (1+r)^{3/2}(l_2^2+r)^{3/2}
(l_3^2+r)^{1/2}}
\eeq
\beq
X_2=l_2l_3\int_0^{\infty}{rdr\over (1+r)^{3/2}(l_2^2+r)^{5/2}
(l_3^2+r)^{1/2}}
\eeq
\beq
X_3=l_2l_3\int_0^{\infty}{rdr\over (1+r)^{3/2}(l_2^2+r)^{3/2}
(l_3^2+r)^{3/2}}
\eeq
\beq
X_4=l_2l_3\int_0^{\infty}{rdr\over (1+r)^{1/2}(l_2^2+r)^{3/2}
(l_3^2+r)^{3/2}}
\eeq
\beq
X_5=l_2l_3\int_0^{\infty}{rdr\over (1+r)^{1/2}(l_2^2+r)^{1/2}
(l_3^2+r)^{5/2}}.
\eeq
For the numerical integration, as in Chandrasekhar (1970ab), we adopt an 
initial equilibrium configuration given by
$l_2=R_2/R_1=0.43223$, $l_3=R_3/R_1=0.34506$, $2X_1=\Omega_*^2/\pi G\rho_*=
0.28403$, which corresponds to the stable 
equilibrium configuration on the sequence of the Jacobi ellipsoids.  
>From the time evolution, the stellar rotation can be read  using 
$\Omega_*/(\pi G \rho_*)^{1/2}=(2X_1)^{1/2}$.

The result of the integration, shown in Fig. 3, 
is similar to that of Chandrasekhar
(1970ab) except for the overall evolution time scale. 
We summarize the main result as follows:
(i) The rotation of the pulsar changes from $\omega_*=0.5329$
(initial) to $\omega_*=0.612$ (final) at which point the pulsar has
transformed into the Maclaurin spheroids (Fig. 3(a)).
(ii) The equatorial area of the pulsar is 
$\pi R_1 R_2=\pi{\bar R}^2(l_2/l_3^2)^{1/3}$. The ratio between the
initial and final equatorial areas is $1.537/1.433=1.073$ reflecting
the deformation due to the change of pulsar rotation.
(iii) The 3-axis eccentricity,
$e_{13}=(1-l_3^2)^{1/2}$, changes from 0.9386 (initial) to 0.8127 (final).
(iv) The equatorial eccentricity, $e_{12}=(1-l_2^2)^{1/2}$, decreases from
0.9018 toward $e=0$ continuously (Fig. 3(c)).
(v) The gravitational radiation luminosity decreases roughly on a time
scale $\sim O(100)t_G$ as expected (Fig. 3(d)).
The evolution of the Jacobi ellipsoid, due to the emission
of the gravitational radiation, spins-up the pulsar as was noted by
Chandrasekhar (1970ab). The spin-up time scale is $\sim O(10^2)t_G$ which is 
similar to $t_{GR}$ in eq. (1-5). 

Despite the spin-up of the pulsar, the electromagnetic emission is not
expected to be simply described by eq. (4-1) due to the change in
geometric configuration and the moment of inertia of the pulsar.
We now consider the evolution of the electromagnetic emission 
during spin-up.
In accordance with our assumption of the flux-freezing, we consider the
pulsar magnetic field as a dipole field described  by
\beq
{\bf B}({\bf R})={3{\bf R}({\bf R}\cdot{\bf \mu_*})\over R^5}-{{\bf \mu_*}
\over R^3}.
\eeq
The total flux passing through a closed surface enclosing the pulsar
vanishes (i.e. ${\bf\nabla}\cdot{\bf B}=0$). For a chosen rotation
axis, 3-axis, we assume that in the comoving frame the magnetic moment
of the pulsar lies on the plane defined by the 3-axis and the 1-axis
and hence ${\bf\mu_*}=(\mu_{*,1}, 0, \mu_{*,3})$. 
The conservation of the magnetic flux through the $23$-plane and
$\nabla\cdot{\bf B}=0$ (or the vanishing total flux for the closed surface 
around the pulsar) combine to require that $\mu_{*,1}$ vary as
(cf. Finn \& Shapiro 1990)
\beq
\mu_{*,1}\sim {R_*l_3\over E(\sqrt{1-(l_3/l_2)^2})}= {\bar R}
\left(l_3^2\over l_2\right)^{1/3}{1\over E(\sqrt{1-(l_3/l_2)^2})}
\eeq
where $R_*={\bar R}/(l_2l_3)^{1/3}$ and $E(x)$ is the complete elliptic
integral of second kind (Abramowicz \& Stegun 1964).
Since the electromagnetic emission from the dipole is given by
\beq
L_{EM}\propto \mu_{*,1}^2\Omega_*^4
\eeq
we can evaluate how the electromagnetic power varies as the pulsar
spins up. 
Roughly, we expect $L_{EM}\propto R_ *^2\Omega_*^4$
assuming the magnetic field is a fossil 
subject to flux-freezing. Thus, the dipole power behavior depends
on whether or not the angular velocity increases faster or slower than
$R_*^2$.

Fig. 3(b) shows the resulting exact behavior of the
electromagnetic power during the evolution driven by the gravitational
radiation emission. The electromagnetic power {\it increases} roughly by
a factor $\sim 2$ which is due to the spin-up and the change of the
magnetic moment.  
That is, during the period of the gravitational radiation emission,
despite the loss of angular momentum, the spin-up of the MSP occurs
and $L_{EM}$ remains essentially constant. Therefore, even in this
stage, the short time scale $t_{GR}$ is not relevant for the characteristic
time scale of the GRB durations. It is the time scale $t_{EM}$,
which is generally much longer than $t_{GR}$, which determines the
intrinsic emission time scale.
If real MSPs differ seriously from incompressible
homogeneous ellipsoids, our results would not be applicable.  It would then
be difficult to assess the likelihood of spin-up
versus spin-down during the period of gravitational radiation, and thus
whether or not $L_{EM}$ increases or decreases in this period.
Another complicated issue is the dynamical instability of the rapidly spinning
neutron stars right after AIC. The new born MSPs could 
be dynamically stable and
secularly unstable only if the rotation rates before AIC lie in a  narrow
region (Chandrasekhar 1969, Shapiro \& Teukolsky 1983). For a wide range
of pre-AIC parameter space, the new born stars are likely to be dynamically
unstable (Durisen et al. 1986, Williams \& Tohlin 1988, Houser et al. 1994).
Such dynamically unstable configurations would lose mass and angular momentum
on the dynamical time scale. Expansion of the mass shed and shock dissipation 
would result in a nearly axisymmetric equilibrium state which is secularly 
unstable
(either due to gravitational radiation or fluid viscosity) and evolves into
the non-axisymmetric configuration on the dissipation time scale. 
More sophisticated simulations will ultimately be useful in  sorting
out these complications.

Given the lack of any alternative realistic calculations, 
we apply the ideal homogeneous ellipsoid MSP model to the discussions on GRBs.
Based on the result that the electromagnetic luminosity does not change 
significantly on a time scale $\sim t_{GR}$, we arrive at the following 
picture for the rapidly spinning pulsars from AIC.
Initially if the AIC-formed pulsar rotates faster than the critical rate,
the angular momentum is lost on the gravitational time scale
\beq
t_{GR}\sim [3\times 10^{-3} s]\epsilon^{-2}\left(\Omega_*\over 10^4s^{-1}
\right)^{-4},
\eeq
and the pulsar spins up. During this stage, newly fromed MSPs could be
identified as progenitors of GRBs through detection of gravitational radiation.
As the equatorial eccentricity $e_{12}\rightarrow 0$, the evolution of the 
electromagnetic power occurs on the typical electromagnetic time scale
\beq
t_{EM}\sim [5\times 10^8s]\left(\Omega_*\over10^4 s^{-1}\right)^{-2}.
\eeq
For pulsars with initial spin rates below the critical rate, the
initial phase is missing and the overall electromagnetic emission time
scale is given by $t_{EM}$.  As pointed out in 
Blackman et al. (1996) (see below), the observed short duration of GRBs 
and the large apparent luminosities would then require that  the emission 
is sharply beamed in a jet which precesses past the line of sight 
to account for the short time scales.

The two classes of MSPs (distinguished by whether the spin rates at birth
are greater or less than the critical rate) 
could be observationally distinguished by the detection
of the gravitational radiation. In the faster class, the initial
phase would be characterized by the emission of the gravitational
radiation.  This is absent in the slower class.

\subsection{Precession of Gamma-ray Jets}
The overall energy budget requires relativistic beaming (eq. (4-5)). 
If their energy source is electromagnetic radiation, the beamed
GRB jets 
may sustain their luminosities for time scales longer than the observed
burst durations.
This is not inconsistent with observations because these jets
will precess due to gravitational interaction with their binary partners. 
Jet precession is a second important feature of the present cosmological
GRB scenario as it accounts for the fact that the observed durations
of the bursts are short, even though the jet lifetime in the precessing frame
can be long (cf. eq. (4-37)).

Blackman et al.  (1996) have pointed out the three 
important precession frequencies in
the pulsar binary systems (Thorne et al. 1986): 
The first is the Newtonian tidal torque 
($\Omega_{T}$). The second is the interaction between the pulsar spin and
the gravitomagnetic field associated with the secondary star's spin 
($\Omega_{J}$). The third ($\Omega_{G}$)
comes from the pulsar spin interaction with the
gravitomagnetic field associated with the secondary's orbital motion and
the secondary's gravitational field, the space-curvature precession, and
the spin-orbit precession from the gravitomagnetic field induced by the
orbital motion of the pulsar in the secondary star's gravitational field. 
The three frequencies, for a maximally spinning 
neutron star and
companion are (Blackman et al. 1996, Thorne et al. 1986)
\beq
\Omega_G\sim [2\times 10^{-8}s^{-1}]
\left[3M_2+(M_*M_2)/(M_*+M_2)\over 3.6M_{\sun}\right]
\left(\Omega_{bin}\over 1.6\times 10^{-3}s^{-1}\right)
\left(D\over 5\times 10^{10} cm\right)^{-1}
\eeq
\beq
\Omega_J\sim 0.5\Omega_G\left(R_2\over 3\times 10^{10}cm\right)^{1/2}
\left(D\over 5\times 10^{10}cm\right)^{-1/2}
\eeq
\beq
\Omega_T\sim 10^{-3}\Omega_G\left(M_*\over 1.4M_{\sun}\right)^{1/2}
\left(D\over 5\times 10^{10}cm\right)^{-1/2}
\eeq
where $\Omega_{bin}$ is the binary orbital frequency.
For a sharply beamed jet, the first precession time scale above is sufficiently
short enough to move the jet out of the observer's line of sight
on a very short precession time scale, while the third precession mode
would keep the jet from returning to the line of sight on $>100$yr time scales.
The observed durations (Meegan et al. 1994) could be
short ($\sim 1$s) even when the intrinsic emission time scale $\sim O(10)yr$
is much longer and the absence of repeaters would be accounted for by
the third precession mode (Blackman et al. 1996).

In the present picture, the source of emission is the electromagnetic dipole
emission. For a wide range of parameters, the precession would be required
if the emission is beamed. The relativistic beaming is naturally expected
in the cosmological scenario. Even when the precession is not required
due to intrinsically short emission lifetime (cf. eq. (1-8)), 
the precession is a natural outcome of binary accretion systems. 
The precession always tends to give
apparent short durations and to ensure absence of repeaters.

The apparent duration is determined by the first precession mode (eq. (4-37))
\beq
\tau_{dur}\sim (\Gamma{\Omega_G})^{-1},
\eeq
where $\Gamma$ is the bulk Lorentz factor for outflowing gamma-ray
emitting blobs. The repetition time is constrained to satisfy (eq. (4-40))
\beq
\tau_{rep}\sim 2\pi\Omega_T^{-1}> 10^9{\rm sec}, 
\eeq
because of the absence of repeaters. We can combine these two relations
with (4-38) and (4-40) to obtain a lower limit on the binary
radius and a lower limit on $\Gamma$.
Using the above parameters for the mass scalings, these limits are given by 
\beq
D\ge7.3\times 10^9\ cm
\eeq
and
\beq
\Gamma_{min}\ge 4 \times 10^5 (\tau_{dur}/1s)^{-1}.
\eeq
In the next section we describe a process which gives such a Lorentz factor.
(For SS433, Martin \& Rees, 1979 used the observation of 1 precession period 
to constrain the binary companion star.)

\subsection{Synchrotron-Inverse Compton Mechanism for Gamma-Ray Emission}
The details of the gamma-ray emission mechanism still remains an outstanding
issue. As a direct application of the derived spin-field correlation,
we consider a particular gamma-ray emission mechanism outlined in
Blackman et al. (1996). 
The derived spin-field correlation for the
AIC pulsars limits the available field strength for a given spin rate,
which could set an interesting constraint on gamma-ray emission mechanisms 
relevant
for MSPs.

We consider the synchrotron-inverse
Compton mechanism (Asseo et al. 1978). 
The charged particles (electron-positron pair plasma)
are accelerated to large $\Gamma$'s (which is essential to avoid the
run-away pair production [Krolik \& Pier 1991, Yi 1993] that would make
the emitting material optically thick) 
by large amplitude electromagnetic waves propagating outside the radius 
(Usov 1994)
\beq
R_{ff}\sim [10^{12}cm]\left(B_*\over 10^{12}G\right)^{1/2}
\left(\Omega_*\over 10^4s^{-1}\right)^{1/2}
\eeq
where the flux-freezing and force-free conditions are broken as the charge 
density decreases below that which can sustain the Goldreich-Julian density
(Usov 1994, Goldreich \& Julian 1969).
At $R=R_{ff}$, the number density of the pair plasma becomes
\beq
N_{ff}\sim [10^6cm^{-3}]\left(R_*\over 10^6cm\right)\left(B_*\over 10^{12}G
\right)^{1/2}\left(\Omega_*\over 10^4 s^{-1}\right)^{5/2}
\eeq
The electron acceleration parameter $\sigma_{ff}$ at this radius becomes
\beq
\sigma_{ff}={L_{EM}\over m_ec^2{\dot F}_{ff}}\approx [5\times 10^7]
\left(R_*\over 10^6 cm\right)^5\left(B_*\over 10^{12}G\right)^{1/2}
\left(\Omega_*\over 10^4s^{-1}\right)^{1/2}
\eeq
where ${\dot F}_{ff}=4\pi R_{ff}^2 c N_{ff}$ is the number flux at
$R=R_{ff}$ (Michel 1984, Usov 1994). 
Solving the equations of motion for a particle in a pulsar wind zone
subject to electromagnetic forces,  leads to the result that 
relativistic electromagnetic 
waves of frequency $\Omega_*$ can accelerate pair plasma to
\beq
\Gamma_{max}\sim \sigma_{ff}^{2/3}\sim 10^5\left(R_*\over 10^6cm\right)^{10/3}
\left(B_*\over 10^{12}G\right)^{1/3}
\left(\Omega_*\over 10^4 s^{-1}\right)^{1/3}
\eeq
(Asseo et al. 1978, Michel 1984).
As mentioned earlier, the emission is beamed into an angle
$\sim \Gamma_{max}^{-2}$ and the correspondingly large luminosities received
per solid angle may be the result 
of this relativistic beaming (Yi 1993, Usov 1994, Blackman et al. 1996). 

As in synchrotron radiation (e.g. Rybicki and Lightman, 1979),
the characteristic emitted frequency of the optically thin
synchro-Compton radiation 
is proportional to  $\Gamma_{max}^3$ or $\sigma_{ff}^2$ from (4-48). 
In particular (Asseo et al. 1978)
\beq
\nu_{sc}\sim [10^{20}s^{-1}]\left(\Omega_*\over 10^4 s^{-1}\right)
\left(\sigma_{ff}\over 10^8\right)^2.
\eeq
The synchrotron-Compton scattering spectrum's tail can extend up to
\beq
\nu_{sc,max}\sim \sigma_{ff}^{4/3}[eB_{ff}/m_ec]\sim [10^{24}s^{-1}]
\left(B_*\over 10^{12}G\right)^{7/6}\left(R_*\over 10^6cm\right)^{23/3}
\left(\Omega_*\over 10^4s^{-1}\right)^{1/6}
\eeq
for a dipole-like field which at $R=R_{ff}$ has $B(R=R_{ff})=B_{ff}$.

Observable gamma-ray emission from the synchrotron-inverse Compton 
scattering mechanism requires
\beq
\left(\Omega_*\over 10^4 s^{-1}\right)^2\left(R_*\over 10^6 cm\right)^{10}
\left(B_*\over 10^{11}G\right)\ge 40
\eeq
or for $R_*=10^6cm$
\beq
\left(\Omega_*\over 10^4s^{-1}\right)\ge 6.3\left(B_*\over 10^{11}G\right)
^{-1/2}
\eeq
which can be compared with the spin-field correlations derived earlier.

Fig. 4. schematically shows the two spin-field correlation lines for the two 
magnetized disk models and the gamma-ray emission line derived above.
Phenomenologically, 
the two models differ only in the effective strength of the field in the
accretion disk $\delta$. That is, for $\alpha=0.1$ and $\gamma=1$ or
$\delta\sim 3$ and for $\gamma_{max}=1$ or $\delta=1$ and the difference
between the two models is mainly due to the effective field strength $\delta$. 
Given the fact that $\delta>1$ is
most likely, the condition for the gamma-ray emission results in a quite
stringent constraint on the spin-field values. It is very likely that 
gamma-ray emitting pulsars from AIC will have $B_*<$ a few $\times 10^{11}G$ 
and $\Omega_*>10^4 s^{-1}$ which puts most of the gamma-ray MSPs in a regime 
close to that of the fizzlers. However, taking into account the uncertainties in
$\delta$, it is likely that there are a small number of gamma-ray MSPs with
$B_*\sim$ a few $\times 10^{11}G$ and $\Omega_*\sim 10^4 s^{-1}$ with 
gamma-ray emission. Any realistic pulsar gamma-ray emission mechanisms are 
expected to be constrained in a similar fashion.

\subsection{Very Strongly Magnetized Pulsars and Observed Gamma-Ray Bursts}

If the gamma-ray emission is isotropic and burst sources are cosmological,
$B_*>10^{15}$G dipole fields
are required (eqs. (1-3),(1-7)). Since the flux-frozen fossil 
field is limited below $\sim 10^{12}$G in MSPs formed by AIC, strong fields
$>10^{15}$G would have to be generated in situ. 

A traditional $\alpha-\Omega$ dynamo can in principle
generate mean fields exponentially through
a combination of helical turbulence, 
differential rotation, and turbulent diffusion (e.g. Parker, 1979;
Moffatt, 1978).  Duncan \& Thomson (1992) suggest that 
for young MSPs, the available turbulent energy and shear motions,
which allow wrapping of field lines around the star,  
can generate dipole fields of order $10^{15}$G.  

We will see
below why this field strength may still be too small, but even so, there
are other challenges to invoking in situ field generation in MSPs:
In the Galaxy for example, exponential dynamo growth of
magnetic field is  essential because of the long  growth
time (e.g. Ruzmaikin et al., 1988) relative to the total available
Galactic lifetime.  
However in the case of MSPs, even if the full feedback leading to exponential
mean field dynamo growth were inoperative and only linear growth 
by differential rotation (e.g. Ruzmaikin et al., 1988) were present, 
the short rotation periods
would imply that strong dipole fields should still be
generated if the shear energy were available
to stretch the field lines by flux freezing.

However, it is precisely the ease with which this could
be expected to occur within the assumption of flux freezing
that might pose a challenge.  If the differential rotation
energy of MSPs (or sub-millisecond pulasars) could be converted into magnetic 
fields (Duncan \& Thompson, 1992), 
then most MSPs would  have $\sim 10^{15}$ Gauss 
fields, or no MSPs should be observed:
If MSPs are formed at birth, then these objects should spin 
down rapidly ($\sim$ 1hr) and would no longer be seen
as MSPs. If the objects are formed later by spinning up, 
then the dynamo should generate strong fields at this stage, and
and here again the objects should spin down.
One difference is that the latter type
of objects would have a crust, which may somewhat keep the field from escaping
as a dipole, but the extent to which this would
keep the rotation from spinning down is not entirely clear.

If MSPs are formed by AIC then the absence of 
such strong fields suggests that
such field growth is not a canonical feature of MSPs.
If there were such AIC formed MSPs, then we would expect to see Galactic 
sources with available dipole power of $>10^{50}$ ergs/sec
without supernova precursors.
Observed field strengths of MSPs are much lower than 
the dynamo field strengths of $\sim 10^{15}$G of Duncan \& Thompson (1992).
In any case, complications with invoking in situ field generation
suggest that constraints on spin-magnetic field relations from
just AIC considerations may be the most important in constraining
MSP properties.

Let us suppose, however, that strong dipole fields 
$>10^{15}$G could be formed in MSPs, (for example by a mechanism
that induces dynamo growth only in newly formed pulsars) 
and address
their consequences in the cosmological scenario. 
A MSP with $\Omega_*\sim 10^4 s^{-1}$ would provide a large enough isotropic 
luminosity $\sim 2\times 10^{51} erg/s$ only when the field strength
\beq
B_*>[10^{16}G] (\xi/0.1)^{-1/2}
\eeq
which depends on the gamma-ray efficiency $\xi$ (e.g. Usov 1994).
Even when $L_{EM}\sim [10^{52} erg/s](\xi/0.1)^{-1}$, the expected
duration of the event is estimated at $t_{EM}\sim 5s$ which would account only
for the longer burst durations only.
This constraint is based on our result from section 4.2
that spin-up (rather than spin-down)
occurs during the period of gravitational radiation, and that 
$L_{EM}$ therefore 
does not decrease on a time scale $t_{GR}\ll 1s$. 
More realistic extensions of the calculations in section 4.2 
are of great interest but they are beyond the scope of the present 
investigation.
This implies that short bursts with $\tau_{dur}\sim 0.1s$ would require
$B_*\sim 10^{17}G$ dipole fields, which is highly unlikely.
Thus in order to account for both cosmological
luminosities and short durations, the field strength has to be close to
$B_*>10^{16}$G. On the other hand, if gamma-ray emission is relativistically
beamed with $\Gamma>10$ as often seen in extragalactic jets, 
$B_*\sim 10^{15}(\Gamma/10)^{-1}$G would be sufficient to account for the
observed luminosity. In this case, the overall duration would become
as long as $\sim 50s$ (eq. (1-8)) which is certainly too long to account for 
short duration bursts. If $\Gamma\gg 10$, even though precession would make 
the apparent duration short, the observed luminosity would be too high unless
the gamma-ray efficiency $\xi \sim \Gamma^{-2}$ is very low. 
Such a low electromagnetic efficiency (and hence gamma-ray
efficiency) is not surprising in GRB models based on MSPs as most of pulsar
rotational energy is carried away by gravitational radiation.

\section{Summary and Discussions}

We have examined the formation of MSPs from  electromagnetically quiet
AIC of magnetized white dwarfs.
Using magnetized accretion disk models, we have derived a
relation  between the stellar spin and  magnetic field, the latter of  
which is assumed to be flux-frozen during  accretion and collapse.
For the high mass accretion rates required for AIC,
the MSPs are born with a typical field strength
less than $\sim 10^{12}G$ whether or not they reach their equilibrium
spin state during the pre-collapse stage. A significantly different spin-field
relation is expected only when the secondary star is a dwarf star with
a very small binary separation, and then the spin and binary orbits would
be synchronized.  Spin-binary orbit synchronization
would be absolutely required for  formation of the strongly magnetized 
($\sim 10^{15}G$) MSPs but the derived spin-magnetic field correlation 
strongly suggests that rapid spin and very strong magnetic fields are 
mutually exclusive. 

This conclusion remains valid within the range of initial conditions
we adopted, which include an almost entire range of initial white dwarf
spin periods. This is not surprising given the short spin-up time scale
(eq. 3-13) leading to near spin equilibrium. 
Siginifcantly different initial conditions 
(such as white dwarf initial spin and mass) and 
physical conditions (white dwarf magnetic field, mass accretion rate), 
which are still appropriate for AIC, in principle, could affect the spin-field 
correlation. Within our disk-magnetosphere model, however, it is highly
unlikely to have MSPs with $B_*\gg 10^{12}$G. The spin-orbit 
synchronization however remains a possible route to MSPs with very strong 
magnetic fields $\gg 10^{12}$G (Usov 1992). It will be crucial for the 
synchronization model to detect $\sim 10^9$G white dwarf field.
Alternatively, it is conceivable that the disk-magnetosphere interaction is
substantially different from what we have described. In the presently available
alternative models, it has been argued that the spin-up by magnetized
accretion disks is much less efficient than in the present model (e.g.
Shu et al. 1994, Lovelace et al. 1995, Li 1996). In this sense, our spin-field
correlation can be regarded as most favorable to formation of MSPs with
strong flux-frozen magnetic fields and hence as providing an upper limit
($\sim 10^{12}$G) for the MSP frozen field.

Given the required high mass accretion rates for AIC, 
and the absence of the observed $\sim 10^9$G field white dwarfs, 
any ultra-strong dipole magnetic fields $\sim 10^{15}$G MSP are 
extremely difficult to produce by AIC (cf. Usov 1992) and, 
as discussed in section 4., they do not seem to be characteristic features
of Galactic MSPs as would be expected if the in situ field production 
(Duncan \& Thompson, 1992) were a canonical feature of MSPs.
A small fraction of magnetized white dwarfs could have $\sim 10^9$G
although such systems have not been observed. These systems would become
MSPs (via AIC) only when the spin-orbit synchronization occurs prior to AIC.
This is because the disk-magnetosphere interaction would most likely to
give long spin periods (eq. 3-7).
Even if such MSPs existed, the constraints from the observed luminosity and
duration severely limit the available parameter space which essentially
requires $B_*>10^{16}$G for bursts with shortest durations.

The spin-field correlation
puts an upper limit on the intrinsic electromagnetic power which
is far less than the observed luminosities of the cosmological
GRBs if the emission were isotropic. We conclude that the
gamma-ray emission has to be strongly beamed in an AIC scenario. 
The small required electromagnetic power would lead to GRBs  much
longer than observed if it were not for precession of the 
beamed gamma-ray emitting jets. 
The typical binary precession periods could account for the
observed durations (Blackman et al. 1996) and absence of repeaters.
Anisotropic emission, mediated by magnetic fields in an AIC
scenario is consistent with overcoming the baryon contamination problem
thoroughly investigated by Ruffert et al. (1996).

Adopting the synchro-inverse Compton mechanism for the gamma-ray emission
as an example, 
we have applied the condition for detectability of the gamma-ray
emission from MSPs. This mechanism favors gamma-ray emission from pulsars
with relatively weak fields
($\sim 10^{11}G$) but fast spins. These pulsars are likely to have
initial conditions similar to those of the pre-collapse white dwarf
systems which result in the fizzlers. In this sense, the fate of the
fizzlers is quite interesting (Tohlin 1984) as is the exact fate of the
dynamically unstable new born stars from AIC (e.g. Houser et al. 1994).

For rapidly rotating pulsars, whose spin frequencies are larger than the
critical frequency, the initial phase of the emission is dominated
by the gravitational radiation. In the simplified picture of the
homogeneous ellipsoids, during the initial phase, the electromagnetic
emission remains nearly constant or increases despite the loss of the angular
momentum. After the short initial phase $<1s$, the pulsars become
axisymmetric, the gravitational radiation becomes small, and the emission is
dominated by the electromagnetic emission. For initially slowly rotating
pulsars, the initial gravitational radiation dominated phase is absent.
In both cases, the binary precession is crucial for the short 
durations. Hence, apart from the initial phase, the long electromagnetic 
emission phase is the inevitable outcome for the pulsars formed in AIC.
The existence of the initial phase in the MSPs spinning above the critical
rotation frequency could be distinguished by the accompanying gravitational
radiation.

Blackman et al. (1996) estimated the required AIC event frequency for
the GRBs. The observed burst rate $\sim 10^{-6}  yr^{-1}$ per galaxy
requires the source event rate $\sim 10^{-4} yr^{-1}$ per galaxy. 
Though the estimated rate is indeed larger than observed rate because of the
beaming, the discrepancy is not nearly as extreme as it
would be if the jet lasted only a duration time:  The long lifetime
of the emission from any given object means
that the probability for observing that object is not merely the beam angle
but the angle swept by the beam that can pass through the line of sight
during the lifetime of the object.
The production rate
combined with the long life time of the MSPs suggest that there would be 
$\le 10$ jetted objects in a typical galaxy. 
When the jet beams are not oriented toward us, these objects would appear
similar to the normal pulsars. The known $\sim 10^5$ cataclysmic variables
(as progenitors of AIC) in our Galaxy and the $\sim 10^2$ low mass X-ray 
binaries (related to MSPs) then imply that a fraction of AIC events may have
resulted in gamma-ray emitting MSPs with the characteristic intrinsic luminosity
$\sim 10^{42-44} erg/s$. It is interesting to point out that the expected events
from electro-magnetically quiet AICs with the canonical luminosity
$\sim 10^{42-44} erg/s$ (cf. eq. (4-4))
and the time scale $\sim 1-10 yr$ (cf. eq. (4-6)) from new-born
MSPs have not been identified. In our scenario, the observed GRBs 
(with relativistic beaming and precession) are the results of such AIC events.
We have  shown that these objects are not the canonical
result of AIC; they fit into a small but available parameter
space, and would not overpopulate the MSP population.

Only if the dipole magnetic fields were $\ge 10^{15}$G
would  the luminosity of the pulsar be high enough to make
an isotropic GRB emission at a cosmological distance and
these fields would have to be produced by neutron star dynamos, if
these are opertative.

If the gamma-ray emission region is close to the neutron star, the critical
field strength $>4.4\times 10^{13}$G in the gamma-ray emission region could 
produce observable features in gamma-ray spectra through photon splitting
(e.g. Adler 1971, Baier et al. 1996, Adler \& Schubert 1996) 
or photon-magnetic field attenuation (e.g. Ho et al. 1990). 
If the emission region is far away from the neutron star surface, 
however, such spectral signatures are not expected. If the GRBs
are cosmological, the severe requirements due to the gamma-ray transparency
(Krolik \& Pier 1991) and baryon loading problem (Meszaros \& Rees 1996)
point to the gamma-ray emission region well away from the neutron star surface.
The field strength in the gamma-ray emission region is likely to
be lower than the critical field strength, which makes the direct confirmation
of the field strength $>10^{13}$G unlikely.

\acknowledgments
We thank G. Field, P. Kumar, R. Narayan, M. Rees, and E. Vishniac for
many helpful suggestions, discussions and comments during various stages 
of this project and the referee of our previous paper who raised the 
issue of the electromagnetic emission from rapidly spinning pulsars. 
Helpful suggestions and criticisms by the referee, V. Usov, are gratefully 
acknowledged. I. Y. acknowledges support from SUAM Foundation.

\clearpage

\begin{figure}[t]
\centerline{\psfig{figure=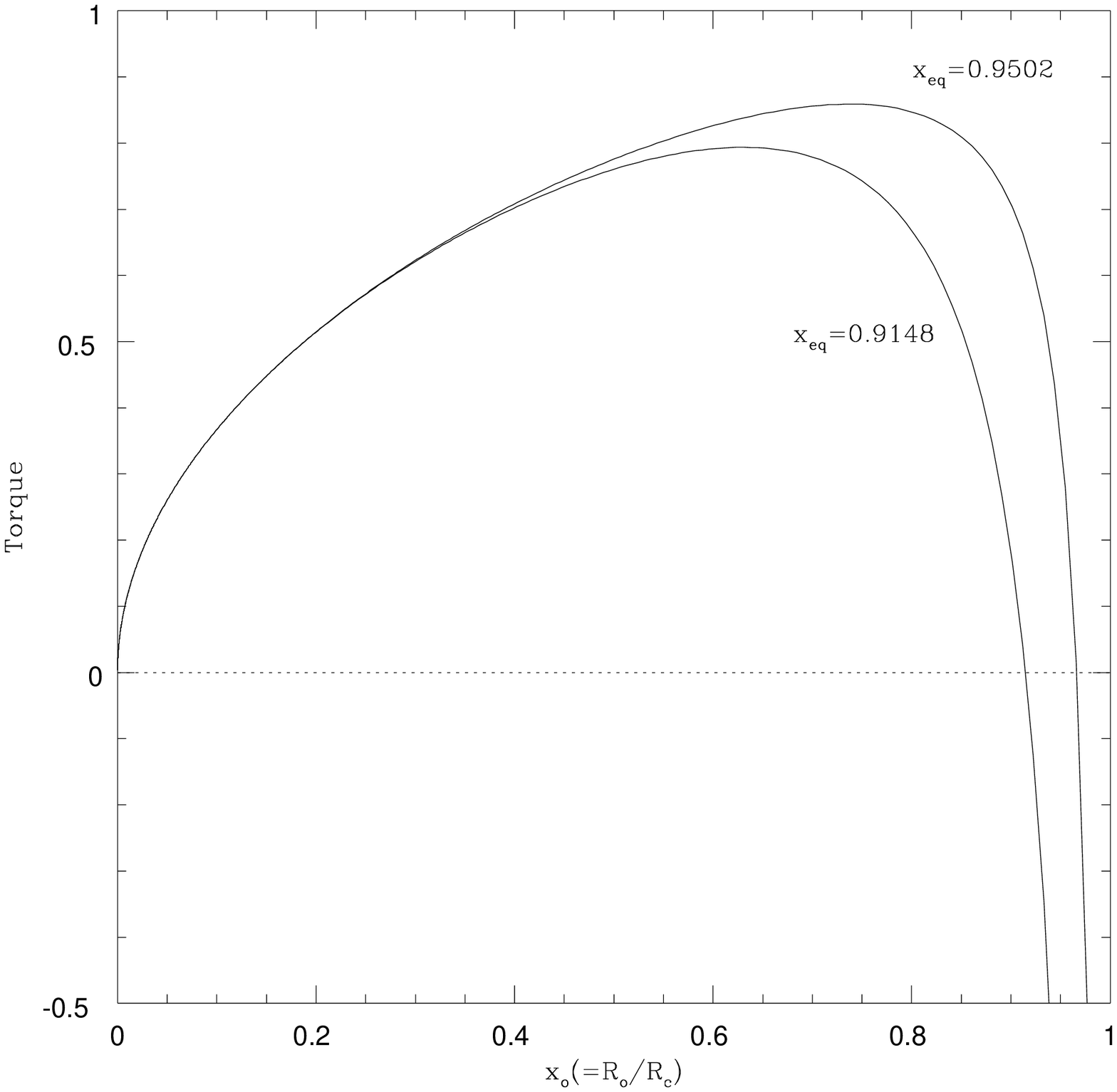,width=14.0cm,height=14.0cm}}
\caption[]
{Torque exerted on the star by the magnetized disk as a function of the
disk truncation radius. The upper curve 
corresponds to the model where the pitch of the azimuthal component of
the magnetic field is limited by the nearly force-free condition in the 
stellar magnetosphere. The lower curve corresponds to the model with the 
diffusive field loss mechanism. The difference is mainly due to the torque 
contribution from the region outside the corotation radius where the two 
models differ. The equilibrium spin is achieved at $x_{eq}=0.9148$ in the 
lower curve and at $x_{eq}=0.9502$ in the upper curve.
}
\end{figure}

\begin{figure}[t]
\centerline{\psfig{figure=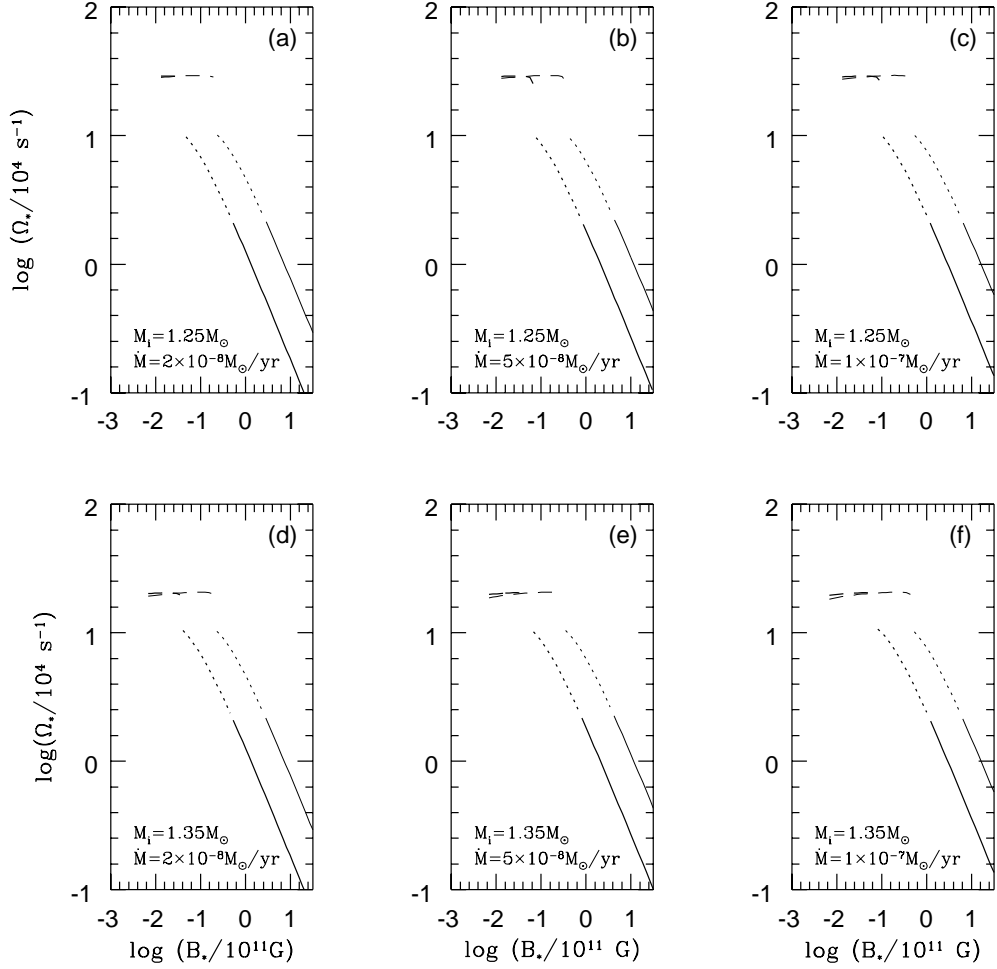,width=14.0cm,height=14.0cm}}
\caption[]
{Spin $-$ magnetic field correlation after collapse from white dwarfs
to neutron stars. 
Initial mass ($M_i$) and mass accretion rate are shown 
in the panels. Initially all white dwarfs are taken as sufficiently
slowly rotating with varying magnetic field strengths.
The thin lines correspond to the pitch-limited magnetic
torque and the thick lines correspond to the model where the field loss
is due to turbulent diffusion. In each panel, the dashed line 
corresponds to the hypothetical critical rotators which are assumed 
to collapse with the conditions exactly at critical rotation. 
The dotted line corresponds
to the fizzlers and the solid line to those directly collapsing to neutron 
stars.
} 
\end{figure}

\begin{figure}[t]
\centerline{\psfig{figure=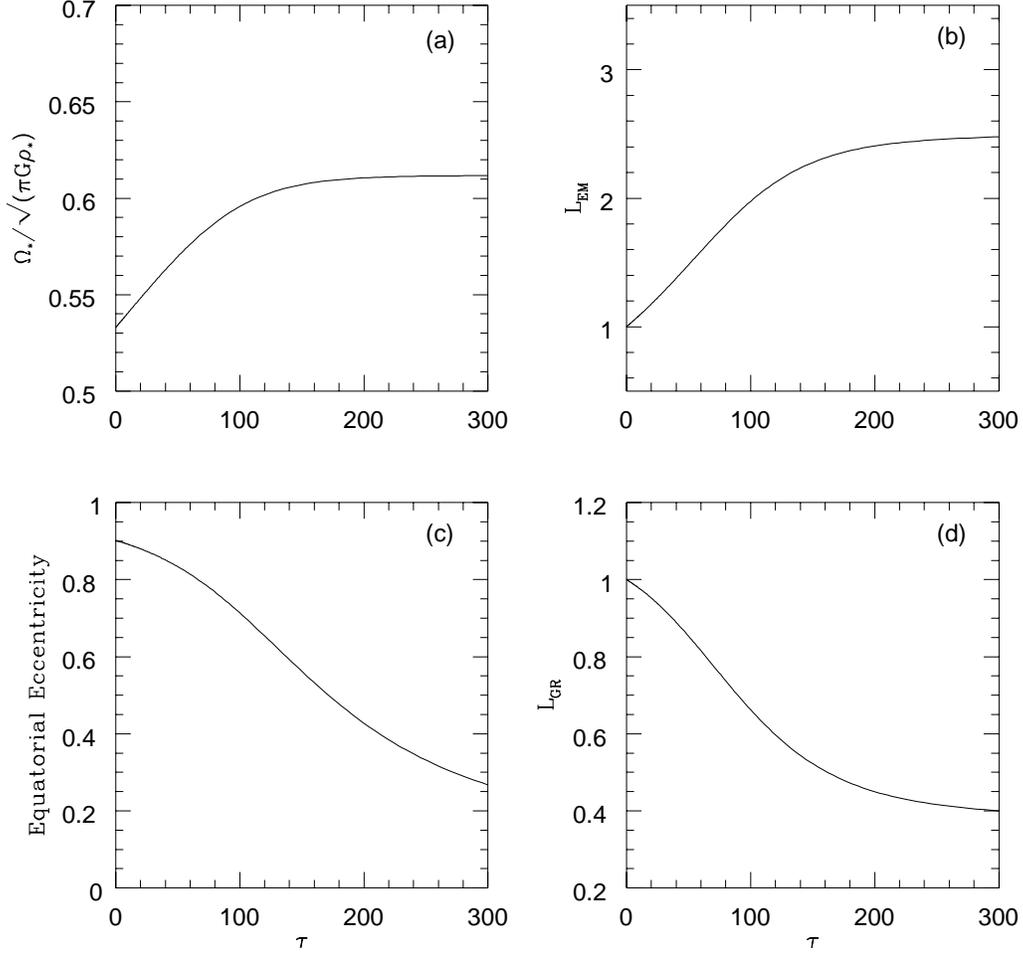,width=14.0cm,height=14.0cm}}
\caption[]
{Time evolution of a rapidly spinning neutron star approximated
as a Jacobi ellipsoid. The panel (a) shows the spin-up of the pulsar 
as a result of the gravitational radiation emission. 
The panel (c) shows the
evolution of the equatorial eccentricity, $e_{12}$, defined in the text. 
The star evolves toward an axisymmetric spheroid with a non-zero $e_{13}$.
The variation of the gravitational radiation luminosity (normalized
by the luminosity at $\tau=0$) is shown in
panel (d). 
The panel (b) is the evolution of the electromagnetic luminosity
(normalized by the luminosity at $\tau=0$)
as a result of the gradual change in the moment of inertia and loss
of angular momentum through the gravitational radiation. 
}
\end{figure}

\begin{figure}[t]
\centerline{\psfig{figure=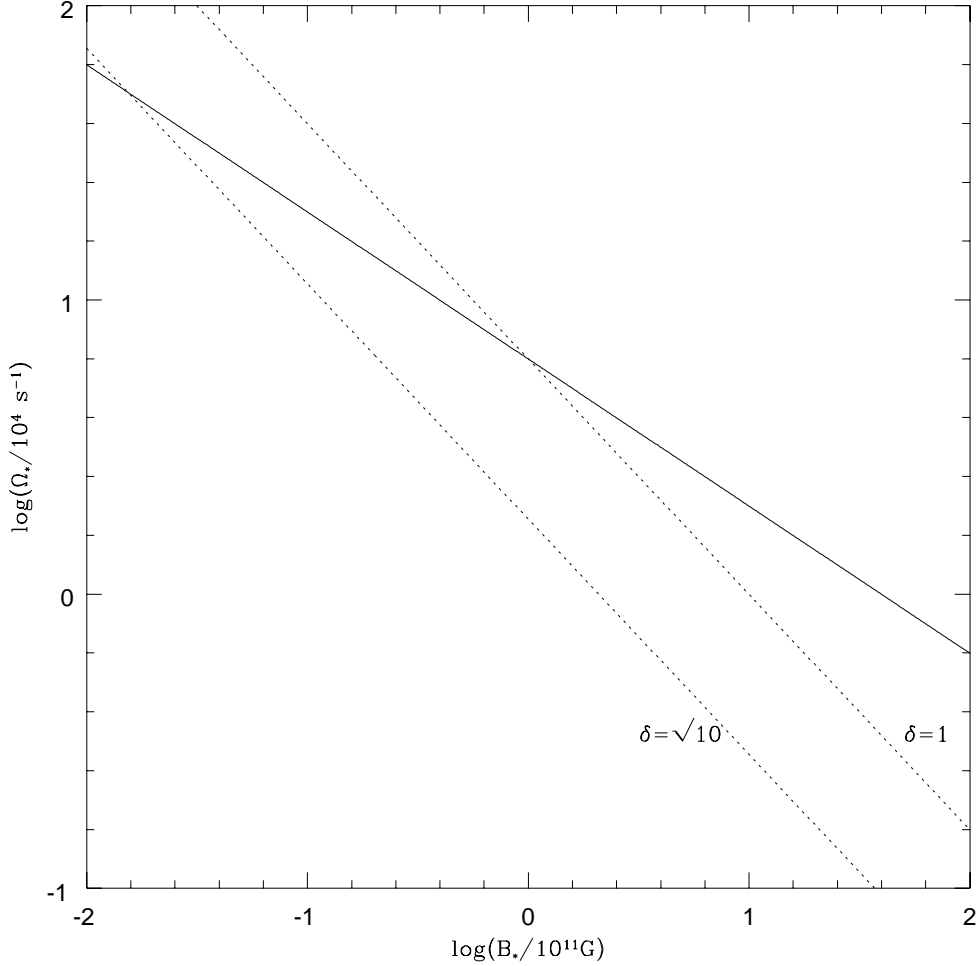,width=14.0cm,height=14.0cm}}
\caption[]
{Schematic constraints for gamma-ray emission from pulsars.
The gamma-ray emission from the synchrotron-inverse Compton scattering 
mechanism is possible when the parameters are above the solid line i.e. 
the spin has a lower bound defined by the solid line for a given field
strength.  The two dotted lines are the predicted spin-magnetic field 
correlations marked by the $\delta$ values.
The upper dotted line is for the pitch-limited model with $\delta=1$ 
and the lower dotted line is for the diffusive loss model with $\delta=3$. 
}
\end{figure}

\end{document}